\documentclass[a4paper,fleqn,usenatbib]{mnras}

\usepackage[T1]{fontenc}
\usepackage{ae,aecompl}
\usepackage{graphicx}
\usepackage{amsmath}
\usepackage{amssymb}
\usepackage{subcaption}
\usepackage{bm}
\usepackage{times}
\usepackage{enumitem}
\usepackage{lastpage}
\usepackage{CJK}

\begin{document}
\begin{CJK}{UTF8}{gbsn}

\newcommand{\gsim}{\,\lower.7ex\hbox{$\;\stackrel{\textstyle>}{\sim}\;$}}
\newcommand{\lsim}{\,\lower.7ex\hbox{$\;\stackrel{\textstyle<}{\sim}\;$}}

\newcommand{\del}[1]{{\color{red}\sout{#1}}}
\newcommand{\add}[1]{{\color{blue}{#1}}}
\newcommand{\delmath}[1]{{\color{red}{\ifmmode\text{\sout{\ensuremath{#1}}}\else\sout{#1}\fi}}}

\title[Efficiency of $B$-field amplification at shocks] 
{The efficiency of magnetic field amplification at shocks by turbulence}

\author[S. Ji et al.]
{Suoqing Ji (季索清),$^1$\thanks{E-mail: suoqing@physics.ucsb.edu} S. Peng Oh,$^{1}$ M. Ruszkowski$^{2,3}$ and M. Markevitch$^{4}$ \\
$^{1}$ Department of Physics, University of California, Santa Barbara, CA 93106, USA.\\
$^{2}$ Department of Astronomy, University of Michigan, 500 Church Street, Ann Arbor, MI 48109, USA.\\
$^{3}$ Department of Astronomy, University of Maryland, College Park, MD 20742 \\
$^{4}$ Astrophysics Science Division, X-ray Astrophysics Laboratory, Code 662, NASA/Goddard Space Flight Center, Greenbelt, MD 20771, USA.}

\setlength{\topmargin}{-0.6cm}
\date{Accepted 2016 September 11. Received September 8; in original form 2016 March 25}

\pagerange{\pageref{firstpage}--\pageref{LastPage}}
\pubyear{2016}

\label{firstpage}
\maketitle

\begin{abstract}
Turbulent dynamo field amplification has often been invoked to explain the strong field strengths in thin rims in supernova shocks ($\sim 100 \, \mu$G) and in radio relics in galaxy clusters ($\sim \mu$G). We present high resolution MHD simulations of the interaction between pre-shock turbulence, clumping and shocks, to quantify the conditions under which turbulent dynamo amplification can be significant. We demonstrate numerically converged field amplification which scales with Alfv\'en Mach number, $B/B_0 \propto {\mathcal M}_{\rm A}$, up to ${\mathcal M}_{\rm A} \sim 150$. This implies that the post-shock field strength is relatively independent of the seed field. Amplification is dominated by compression at low ${\mathcal M}_{\rm A}$, and stretching (turbulent amplification) at high ${\mathcal M}_{\rm A}$. For high $\mathcal{M}_{\rm A}$, the $B$-field grows exponentially and saturates at equipartition with turbulence, while the vorticity jumps sharply at the shock and subsequently decays; the resulting field is orientated predominately along the shock normal (an effect only apparent in 3D and not 2D). This agrees with the radial field bias seen in supernova remnants. By contrast, for low $\mathcal{M}_{\rm A}$, field amplification is mostly compressional, relatively modest, and results in a predominantly perpendicular field. The latter is consistent with the polarization seen in radio relics. Our results are relatively robust to the assumed level of gas clumping. Our results imply that the turbulent dynamo may be important for supernovae, but is only consistent with the field strength, and not geometry, for cluster radio relics. For the latter, this implies strong pre-existing $B$-fields in the ambient cluster outskirts.
\end{abstract}

\begin{keywords}
elementary particles -- magnetic fields -- radiation mechanisms: non-thermal -- cosmic rays -- galaxies: clusters: general.
\end{keywords}

\section{Introduction}
\label{section:intro}

Strong magnetic fields are often inferred from synchrotron emission immediately downstream of shocks, often far in excess of what might be expected in the ambient environment. Two key examples are radio relics in galaxy clusters and supernova (SN) thin rims in the interstellar medium (ISM). The origin of these strong magnetic fields is still debated. In this paper, we carefully assess the possibility that fields are amplified by turbulence in the postshock region. 

Radio relics are diffuse, large-scale ($\sim$ Mpc) sites of synchrotron emission on the periphery of galaxy clusters undergoing mergers (e.g., see \citet{ferrari08,bruggen12,brunetti14} for reviews). They are generally thought to be associated with shocks, due to their elongated morphology, the fact that they often come in pairs aligned with the merger axis, and the coincidence of X-ray detected shocks with radio relics. They are also polarized (usually $\sim 10-30\%$), with $B$-fields aligned with the shock front. Mach numbers inferred from X-ray observations and radio spectral indices indicate weak shocks ($\mathcal{M} \sim 2-4$), consistent with expectations from cosmological simulations \citep{hoeft08,battaglia09,skillman11}. Radio relics offer an opportunity to study particle acceleration and magnetic field amplification in a large-scale, low Mach number regime far removed from the high Mach number conditions probed by local supernova remnants (SNRs). A spectacular example is CIZA J2242.8+5301, the
`sausage relic' \citep{van-weeren10}, a large ($\sim 2$ Mpc long;
located $\sim 1.5$ Mpc from the cluster center) double radio relic
system. The post-shock radio spectral index was used to infer the
particle spectral slope and hence the shock compression ratio and Mach
number ($\mathcal{M} \sim 4.6$)\footnote{Note, however, that for this relic, a shock has not been detected in X-rays.}, while the increase in the spectral index
(from $\sim 0.6$ to $\sim 2.0$ across the narrow $\sim 55$
kpc relic width) toward the cluster center---spectral aging due to
synchrotron and inverse Compton losses---was used to infer magnetic
field strengths of $\sim 5 \, \mu$G. Strong ($\sim 50-60 \%$)
polarization was observed.

The origin and orientation of the high $B$-fields in such relics presents an interesting puzzle. Faraday rotation measurements near the cluster center indicate $\mu$G fields, generally thought to be due to the action of a turbulent dynamo\footnote{Note, however, that a number of competing explanations for magnetogenesis exist, ranging from AGN injection \citep{furlanetto01} to small scale plasma instabilities \citep{schekochihin05}.} which brings the $B$-field up to equipartition with the turbulent velocity field \citep{subramanian06,dubois08,dolag09}. However, the existence of fields of similar $\sim \mu$G strength at distances approaching the cluster virial radius is harder to explain. 

Supernova remnants also show amplified magnetic fields. In some regions of the remnant, they can be up to $\sim 100$ times stronger than present in the ambient ISM, far greater than the compressional amplification of $\sim 4$ possible at a strong shock. Particularly interesting are the detection of synchrotron thin rims in young SN remnants where non-thermal emission extends to the X-ray, where $\sim 50-100 \, \mu$G fields are inferred immediately downstream of the shock (e.g., \citet{bamba05}). The thin widths are explained by rapid electron cooling, entirely analogous to the radio relic case\footnote{The thin width could occur if magnetic fields decay rapidly \citep{pohl05}, though this can be ruled out in some cases (e.g., SN1006; \citet{ressler2014magnetic}), since the rim width should be energy-independent.}. Amplification of pre-shock fields must be extremely rapid due to the short time available before fluid is advected downstream. Such strong fields are also required to confine and accelerate cosmic rays (CRs) up to the observed `knee' at $\sim 10^{15}$eV. Farther downstream, even higher fields might be possible \citep{vink03}. X-ray emission is filamentary and rapidly time variable, and in some cases fields as high as 1 mG have been suggested \citep{uchiyama07}. Polarization is relatively modest, but there appears to be a net radial orientation in the body of the supernova remnant \citep{dickel91,delaney02,reynoso13}. 

There are at least 3 possible explanations for amplification of post-shock magnetic fields: (i) Compressional amplification \citep{iapichino12}. This naturally explains the orientation of the post-shock field in radio relics, since compression only amplifies the component parallel to the shock. However, the amplification is modest ($\sim 2.5$ for a typical shock strength ${\mathcal M} \sim 3$), which implies strong ambient $B$-fields, perhaps generated by turbulence. If this is indeed the only viable mechanism, then relics provide power indirect constraints on $B$-fields and turbulence in cluster outskirts, which are difficult to obtain by other means. (ii) Current-driven instabilities \citep{bell04}, where the return current of thermal particles as CRs drift upstream amplify the traverse $B$-field, and in particle-in-cell (PIC) simulations has been shown to amplify $B$-fields by $\sim 10-45$ \citep{riquelme09,riquelme10}. A major disadvantage is that fields are only generated on length scales comparable to the gyro-radius of the streaming particles. An inverse cascade is thus required to bring these fields to larger scales. Hybrid models where streaming CRs drive a turbulent dynamo also exist \citep{beresnyak09,drury12,bruggen13}. (iii) Turbulent amplification.  Studies of the interaction between shocks and ISM with density inhomogeneities due to Kolmogorov turbulence found a small scale dynamo driven by the baroclinic vorticity generated during shock-cloud interactions, which could indeed amplify the fields significantly \citep{giacalone07,inoue09,guo12,sano12,fraschetti13,inoue13}. Turbulent amplification of $B$-fields due to the Richtmyer-Meshov instability has also been seen in laboratory-produced shock waves \citep{meinecke14}. Small scale dynamos can provide exponential amplification on the eddy turnover time of the smallest eddies \citep{kulsrud05,balsara05}.

In this paper, we critically examine the last possibility, since gas clumping is expected to be omnipresent in the ISM and intracluster medium (ICM). Note that all previous studies have focussed on conditions relevant to supernova remnants. There has been no detailed numerical work on turbulent field amplification applicable to cluster radio relics; we remedy this omission here. We also study conditions appropriate to the supernova case in parallel. We conduct a detailed survey of parameter space, convergence studies, comparison of 2D and 3D simulations, and compare to analytic expectations. The goal is to validate physical expectations and analytic scalings (e.g., scaling of $B$-field amplification with shock Mach number, \citet{fraschetti13}) so that reliable estimates of the importance of turbulent field amplification can be made. The outline of the paper is as follows: in \S\ref{sect:methods}, we describe our computational methods. In \S\ref{sect:theory}, we describe key parameters and theoretical expectations, and in \S\ref{sect:results}, we present our findings. We conclude in \S\ref{sect:conclusions}. Throughout the text, ${\mathcal M}$ refers to sonic Mach number, while ${\mathcal M}_{\rm A}$ refers to Alfv\'en Mach number. 

\section{Methods}
\label{sect:methods} 

The FLASH code \citep{fryxell00} developed by the FLASH Center of the University of Chicago is the main code for our simulations. FLASH uses a directionally unsplit staggered mesh (USM) MHD solver to solve the following fundamental governing equations of inviscid ideal magnetohydrodynamics:
\begin{align}
  \frac{\partial\rho}{\partial t} + \bm{\nabla} \cdot (\rho \bm {v}) &= 0 \\
  \frac{\partial\rho \bm{v}}{\partial t} + \bm{\nabla} \cdot (\rho \bm{v} \bm{v} - \bm{B} \bm{B}) + \bm{\nabla} p_* &= 0 \\
  \frac{\partial \rho E}{\partial t} + \bm{\nabla} \cdot \left[\bm{v} (\rho E + p_*) - \bm{B} (\bm{v} \cdot \bm{B}) \right] &= 0 \\
  \frac{\partial \bm{B}}{\partial t} + \bm{\nabla} \cdot (\bm{v}\bm{B} - \bm{B} \bm{v}) &= 0
\end{align}
where $p_* = p + B^2 / (8 \pi)$ is the total pressure including both gas pressure $p$ and magnetic pressure $B^2 / (8 \pi)$, and $\rho E$ is the total energy density with $E = 1/2 \rho v^2 + \rho \epsilon +  B^2 / (8 \pi)$. The MHD solver is based on a finite-volume, high-order Godunov method combined with a constrained transport (CT) type of scheme which ensures the solenoidal constraint of the magnetic fields on a staggered mesh geometry \citep{tzeferacosetal12, lee2013solution}. The Roe Riemann solver with piecewise parabolic (PPM) spatial reconstruction is adopted in our simulations.

For computational speed, we mostly perform 2D simulations, but also run a suite of 3D simulations and compare and contrast the results. Our fiducial simulations have a resolution of $2048 \times 512$. Unless otherwise noted, all results presented are at this resolution. 

We begin by specifying the initial density field. As discussed in \S\ref{sect:key_parameters}, we parametrize the density field as a lognormal distribution with a Kolmogorov-like power spectrum, as in \citet{giacalone07} (see \S\ref{sect:key_parameters} for justification of these choices). The lognormal density distribution is given by:
  \begin{align}
    \rho(x,y) = \rho_0 \mathrm{exp}(f_0 + \delta f),
  \end{align}
  with
  \begin{align}
    \delta f = 
    \begin{cases}
      \displaystyle \sum_{n=1}^{N} \sqrt{2\pi k_n A \Delta k_n P_\mathrm{2D}(k_n)} \times \mathrm{exp}\left[i(k_x x+ k_y y + \phi_n)\right] \\ 
      \qquad\qquad\qquad\qquad(\text{2D},\ k_x^2+k_y^2 = k_n^2)
     \\
     \displaystyle \sum_{n=1}^{N} \sqrt{2\pi k_n^2 A \Delta k_n P_\mathrm{3D}} \times \mathrm{exp}\left[i(k_x x+ k_y y + k_z z + \phi_n)\right] \\
     \qquad\qquad\qquad\qquad(\text{3D},\ k_x^2+k_y^2+k_z^2 = k_n^2)
    \end{cases}.
  \end{align}
where $N$ is the total number of the modes, $k_x$, $k_y$ and $\phi$ determine the directions and phases of the waves which are randomly chosen for each $n$, $A$ is the normalization constant, and $P(k)$ is power spectrum. To set up a density perturbation field, we adjust $A$ to obtain a certain clumping factor $C \equiv \langle\rho^2\rangle/\langle\rho\rangle^2$, and adjust $f_0$ to obtain a certain $\langle\rho\rangle$.
We adopt a Kolmogorov-like power-law spectrum to model spatial variations of the gas clumping:
\begin{align}
  P_\mathrm{2D}(k) & = \frac{1}{1+(kL)^{8/3}} \\
  P_\mathrm{3D}(k) & = \frac{1}{1+(kL)^{11/3}}
\end{align}
where $L$ is the coherence length. Since the power spectrum flattens for $k^{-1} > L$, this imposes an outer scale to the density fluctuations. For the initial density perturbation, we select the $N = 1000$ wave vectors with wavelength $0.05 \leq \lambda \leq 1$ with the simulation domain size of $20\times 5$, and for the power spectrum we apply $L = 0.5$; this represents an outer scale for the density fluctuations. 

We have also explored various setups of initial magnetic field, which include perpendicular and parallel fields\footnote{Following convention, our designation of perpendicular/parallel are with respect to the shock normal.} as well as an isotropic field (in 3D). To generate the 3D isotropic magnetic field, we ensure it is divergence-free by taking the curl of a magnetic potential $\bm{A}$:
\begin{align}
  \bm{B} = \nabla \times \bm{A}
\end{align}
where the magnetic potential $\bm{A}$ follows the power spectrum: 
\begin{align}
  P_A(k) \propto |\bm{A}|^2 \propto k^{-17/3}
\end{align} 
Since $\bm{B}(k) = i k \times \bm{A}(k)$, this implies a Kolmogorov-like power-law spectrum for the magnetic field:
\begin{align}
  P_B(k) \propto k^2 P_A(k) \propto k^{-11/3}
\end{align}
The Kolmogorov-like scaling assumes that the initial $B$-field has reached saturation with a small-scale dynamo, due to a Kolmogorov-like velocity field. We keep the range of wave vectors $k$ the same as those in density field generation. We generate the perturbation modes in $k$ space and take an inverse fast Fourier transform (FFT) to get the magnetic potential in real space. Finally, we adjust the amplitude of magnetic field to obtain specific values of plasma $\beta$.

To introduce a shock, we set up a rigid reflecting piston with a specific velocity $v_{\rm p}$ moving in the positive $x$ direction. A shock propagating at $v_{\rm s} = r/(r-1) v_{\rm p}$ moves outward from the piston; for a strong shock, $v_{s}=4/3 v_{p}$. In the frame of the piston, this setup appears identical to that simulated by \citet{giacalone07}, where moving fluid slams onto a rigid boundary, with a shock propagating outward. Since the boundary is rigid, the post-shock fluid has no net $x$-velocity relative to the piston, i.e. $\langle v_{x} \rangle = v_{\rm p}$. We implement inflow (outflow) boundary conditions on the left (right) and periodic boundary conditions on the top (bottom) as implemented in the FLASH code, which satisfy constrained transport and maintain zero $B$-field divergence. We adopt code units where $\rho=1, c_{\rm s}=1$ in the unperturbed pre-shock medium. We also implement default FLASH values for explicit artificial viscosity ($0.1$ in code units) for stability purposes. Explicit diffusion significantly reduces mixing and dissipation in grid simulations \citep{lecoanet16}, allowing an effective increase of the inertial range. All the same, we take care to explicitly test convergence properties. In practice, we have found that at our resolutions, we are dominated by numerical viscosity at the grid scale; simulations without artificial viscosity are very similar.
  
\section{Theoretical Expectations} 
\label{sect:theory} 

\subsection{Key Parameters}
\label{sect:key_parameters}

In what follows, we consider $B$-field amplification behind a planar shock impinging on a clumpy medium. We would like to understand the $B$-field profile behind the shock, and in particular the immediate post-shock field, the maximum amplification $B_{\rm max}/B_0$, the length scales on which it reaches this peak, and the post-shock magnetic geometry. We would also like to understand the relation of $B$-field amplification to vorticity generation and amplification. The assumption of plane geometry suffices whenever the shock curvature radius (or the scale on which the shock Mach number varies along the shock surface) is much larger than the scale height of density fluctuations. The latter determines the scale on which the shock surface is deformed and baroclinic vorticity is generated. This is generally satisfied for all ISM and ICM shocks, except for special cases of intersecting shocks, or the intersection of a shock with a large scale filament. 

Before proceeding further, it is worth reminding ourselves of all the variables on which our calculations might depend: 

\begin{enumerate} 

\item \emph{Pre-existing density field.} We parametrize the pre-existing density field as a lognormal distribution with a Kolmogorov-like power spectrum\footnote{We do not expect our results to be very sensitive to spectrum of assumed density fluctuations. For instance, single-mode analysis or RMI $B$-field amplification \citep{sano12}, gives broadly consistent conclusions.}. Lognormal density distributions are observed directly in the ISM \citep{ridge06,wong08} and seen in hydrodynamic simulations of the ICM \citep{zhuravleva13}, and can be understood from the central limit theorem from the multiplicative effect of random, uncorrelated compressions/rarefactions (whose logarithms therefore add to give a normal distribution). While an additional high density tail may exist in conjuncture with the volume-filling lognormal component \citep{federrath10,zhuravleva13}, we shall see that beyond a certain level of clumping, the turbulence induced by shock-clump interactions is only weakly dependent on the clumping factor. Kolmogorov-like\footnote{While Kolmogorov turbulence assumes incompressible hydrodynamic flow, the ISM and the solar wind (where similar scalings are seen -- e.g., \citet{Roberts10}) are magnetized, compressible flows. Goldreich-Sridhar theory \citep{goldreich95} does predict that Alfv\'en and slow modes should have a Kolmogorov-like cascade perpendicular to the $B$-field, for incompressible flow. Naively, a Burgers-like $k^{-2}$ scaling might be expected for supersonic compressible flow \citep{kritsuk07}. Currently, the origin of this Kolmogorov-like scaling is not fully explained \citep{lazarian00,elmegreen04,kevlahan09}; we simply adopt it as an empirical fact.} density power spectra are inferred from ISM scintillation, producing the `big power-law in the sky' \citep{armstrong95} over 11 orders of magnitude, from $10^{7}$ to $10^{18}$cm, and (less directly) in the ICM, from X-ray surface brightness fluctuations \citep{gaspari13a,zhuravleva15}. Magnetic power spectra with Kolmogorov-like scalings are also seen via Faraday rotation in clusters \citep{Vogt05}. These choices introduce the parameters:

\item \emph{$L_{\rm max},L_{\rm min}$.} These are the maximum and minimum scale of density fluctuations. Since MHD is scale free, the outer driving scale $L_{\rm max}$ establishes a characteristic scale; in particular, when turbulent amplification dominates, it establishes the distance downstream (typically $\sim$ few $L_{\rm max}$), on which the $B$-field peaks. This in term sets the required length of the box. In principle, $L_{\rm min}$ establishes the peak post-shock vorticity $\omega \sim u/L_{\rm min}$, and thus the timescale (lengthscale) on which exponential amplification takes place. This is relevant to whether the strong $B$-fields seen in cluster radio relics and supernova thin rims can be amplified sufficiently quickly. In practice, unless $L_{\rm min}$ is large, the post-shock vorticity in our simulations is determined by grid-scale and resolution effects.

\item \emph{$l_{\rm diss}$ (or $\Delta x$).} These are the scales on which turbulent motions or $B$-fields are dissipated. In practice, since we cannot resolve these scales, the dissipation scale is given by either the small amount of explicit viscosity we have implemented, or grid scale effects. Estimating the actual Reynolds number of a numerical MHD flow is a thorny issue which resists simple characterization, and we demur from doing so (for discussions of the non-trivial considerations involved in estimating the Reynolds number of numerical hydrodynamic flow, see \citealt{aspden09,radice15}). In \S\ref{sect:results}, we perform a convergence study to show the critical $L_{\rm max}/\Delta x$ needed to achieve convergence.

\item \emph{$C \equiv \langle \rho^{2} \rangle/\langle \rho \rangle^{2}$.} The clumping factor, or amplitude of density fluctuations. We shall see that once the clumping factor exceeds a fairly small threshold, $C \gsim 1.5$, results are only weakly dependent on its exact value. Such a threshold is comparable to or exceeded by inferred clumping in the ISM \citep{leroy13} and ICM \citep{simionescu11,battaglia13,zhuravleva13}.

\end{enumerate}

Other important parameters include:

\begin{enumerate}

\item \emph{$\mathcal{M}_\mathrm{A}$ ($\mathcal{M}_\mathrm{S}$).} The Alfv\'en (sonic) Mach number of the shock. This determines the amount of bulk kinetic energy which is available to be turned into turbulent kinetic energy, which in turn amplify $B$-fields up to equipartition values, via the turbulent dynamo. We shall find that the Alfv\'en Mach number $\mathcal{M}_{\rm A}$ is the most important parameter for determining amplification via the Richtmyer-Meshov instability (RMI).

\item \emph{$\beta$.} Plasma beta parameter, $\beta \equiv P_{\rm therm}/P_{\rm B}$. This determines the initial importance of magnetic tension and magnetic pressure in modifying the shock, and vorticity/$B$-field amplification. It also sets the relation between the sonic and Alfv\'en Mach number, $\mathcal{M}_{\rm A} \approx \beta^{1/2} \mathcal{M}_{\rm S}$. For the ISM, typically $\beta \sim 1$, while for the ICM, $\beta \sim 50$ in the core, potentially declining to values as low at $\beta \sim$few at the cluster outskirts. The plasma $\beta$ in the ICM outskirts is highly uncertain. However, note that for Coma, Faraday rotation measurements indicate that $B\propto \rho^{\alpha_{\rm B}}$, where $\alpha_{\rm B} \approx 0.3-0.7$ \citep{bonafede10}. Thus, this implies that $\beta \propto \rho^{1-2 \alpha_{\rm B}} T$ could fall by a factor of 10 or so from the cluster core to outskirts, for larger values of $\alpha_{\rm B}$. This is also consistent with the scaling $P_{\rm turb}/P_{\rm therm} \approx 0.2 (r/R_{\rm 500})^{0.8}$ seen in cosmological simulations \citep{shaw10,battaglia12}, which implies that the turbulent energy density (and by extension the $B$-field energy density, if the two are in equipartition) could become comparable to the thermal energy density in the cluster outskirts. We therefore consider both high and low values of $\beta$ for the low Mach number ICM shocks.

\item \emph{Initial field geometry.} We expect $B$-fields in the ICM/ISM to be tangled. We can isolate the effects of initial $B$-field anisotropy by considering parallel and perpendicular fields. $B$-field amplification by stretching (turbulent dynamo) is independent of initial field orientation, since the field quickly becomes tangled, losing memory of initial conditions. However, compressional amplification is a factor $r$ (1) for perpendicular (parallel) fields, where $r$ is the density compression ratio.

\item \emph{Initial velocity field.} Pre-existing turbulent motions in the ICM/ISM can be decomposed into compressive and solenoidal components. Compressive motions result in density fluctuations, which we have modeled. However, we shall ignore the solenoidal component. Pre-existing vorticity is amplified at a shock: just as for the $B$-field, the component parallel to the shock normal is continuous, while the perpendicular component is enhanced by the density compression ratio $r$. This enhanced vorticity $ \omega \sim ({\rm cos}^{2}\theta + r^{2} {\rm sin}^{2}\theta)\omega_{o}$,  for a vortex tube with initial vorticity $\omega_{o}$ inclined at angle $\theta$ to the shock normal, is available to amplify $B$-fields. In practice, pre-existing vorticity is only important for low $\mathcal{M}_{\rm A}$ shocks, before it is overwhelmed by baroclinically generated vorticity. We have verified this theoretical expectation directly in numerical simulations.

\item \emph{2D vs 3D.} Most of our simulations are done in 2D, although we run several 3D simulations to confirm our results. These are critical because MHD turbulence can be significantly different in 2D and 3D. In particular, 2D turbulence has only 1 non-zero component of vorticity. Thus, 2D prevents the stretching of vortex tubes ($\propto {\bm \omega} \cdot \nabla {\bm v}=0$), which plays an essential role in the 3D turbulent cascade. This in turn may impact the turbulent dynamo. For the most part, we find broadly similar results in 2D and 3D, with the notable exception of post-shock field geometry.

\end{enumerate}

\subsection{Vorticity and Magnetic Field Amplification Across Shocks}

There are two key pieces of physics which are worth reviewing before we proceed. One is the jump in vorticity across a rippled shock interacting with density enhancements (the RMI). The subsequent downstream turbulence enables a small-scale dynamo. The second is the time-dependent evolution and saturation of the turbulent dynamo. 

For an MHD shock, the component of vorticity normal to the shock is continuous ($\delta \omega_{\rm b}=0$), and the jump in the component parallel to the shock is \citep{kevlahan97,porter15}:
\begin{eqnarray}
\delta\omega_{z} =  \mu\omega_{z,1} - \mu u_{n,1} \frac{1}{\rho_{1}}\frac{\partial \rho_{1}}{\partial s} \left( \frac{1}{1+\mu} - \frac{c_{\rm s,1}^{2}}{u_{n,1}^{2}} \right) + \frac{\mu^{2}}{1+\mu} \frac{\partial u_{n,1}}{\partial s} \nonumber \\
+ \frac{1}{u_{n,1} \rho_{1}} \left( \frac{\partial P_{\rm B,2}}{\partial s} - (1+ \mu) \frac{\partial P_{\rm B,1}}{\partial s} + \hat{\bm q} \cdot {\bm T}_{2} + (1+ \mu) \hat{\bm q} \cdot {\bm T}_{1} \right),
\end{eqnarray}
where the subscripts $1,2$ refer to the upstream (downstream) fluid, and the shock compression ratio $r=\rho_{2}/\rho_{1}=1+\mu$. We also define $\hat{\bm q}, \hat{\bm z}$ to be orthogonal directions tangent to the shock, $\partial/\partial s = \hat{\bm q} \cdot \nabla$, and ${\bm T}= {\bm B} \cdot \nabla {\bm B}$ is the magnetic tension. The first term reflects vorticity enhancement during shock compression due to conservation of angular momentum (for similar reasons, the shock perpendicular $B$-field is continuous, while the shock parallel $B$-field is enhanced due to flux conservation). The second term, which depends on a density gradient, reflects the baroclinic generation of vorticity across the shock in a non-uniform flow (cast in a more general form than $\nabla P \times \nabla \rho$). The third term accounts for vorticity generation due to variations in the shock normal speed (and hence refraction of streamlines). For instance, vorticity generation due to curved or intersecting shocks falls into this category. The last four terms reflect the Maxwell stresses due to $B$-fields. These are potential sources of vorticity. For instance, magnetic pressure is anisotropic, thus implying that the total pressure is no longer barotropic, and allowing baroclinic vorticity generation. Magnetic tension due to a field with varying angle of inclination to a the shock can also induce a vorticity jump \citep{porter15}, or suppress vorticity \citep{fraschetti13}. However, both of these effects scale as $\sim 1/\mathcal{M}_{\rm A}^{2}$; for the high $\mathcal{M}_{\rm A}$ shocks we consider, they are negligible. For the most part, we can focus solely on baroclinic generation of vorticity, which produces a vorticity jump of $\delta \omega \sim v_{\rm sh}/l_{\rho}$, where $v_{\rm sh}$ is the shock speed and $l_{\rho}$ is the density scale height. 

How does the post-shock solenoidal velocity field amplify $B$-fields? The magnetic field obeys the induction equation which closely parallels that for vorticity. Studying the evolution of the vector field is less useful (since for an isotropic field, $\langle B \rangle = 0$); more illuminating is the equation for the evolution of magnetic energy density:
\begin{align}
  \frac{1}{2}\frac{{\rm d} |\bm{B}|^2}{{\rm d} t}
    = \bm{B} \cdot(\bm{B}\cdot\bm{\nabla})\bm{v}-|\bm{B}|^2 \bm{\nabla}\cdot \bm{v},
\label{eqn:Bfield_amp_terms}
\end{align}
where the left hand side is the Langragian time derivative. 
The first term represents $B$-field amplification by stretching, which is the principal mechanism of the turbulent dynamo. The second represents field amplification by compression. We shall subsequently use this equation to measure the relative importance of stretching and compression. Note that in principle amplification by compression is a reversible process, while stretching is irreversible and represents a true dynamo. 

From equation (\ref{eqn:Bfield_amp_terms}), stretching results in exponential amplification $E_{\rm B} \propto {\rm exp}(t/\tau_{\rm m})$ on a timescale $\tau_{\rm m} \sim l/v \sim \omega^{-1}$, i.e., the inverse of the vorticity generated at the shock. Note that $\omega \propto v/l \propto l^{-2/3}$ for Kolmogorov turbulence; i.e., it is dominated by motions at the smallest scales $l_{\rm m}$. Estimates of turbulent amplification which assume amplification timescales of order the eddy turnover time at the outer scale $t_{\rm edd} \sim L_{\rm max}/v$, sometimes argue that turbulent amplification cannot explain the large post-shock field, which appears almost immediately after the shock.

However, the $B$-field efolds on the eddy turnover time of the {\it inner} scale, with $\omega \sim \tau_{\rm m}^{-1} \gg t_{\rm edd}^{-1}$. The efolding length is thus of order $l_{\rm e} \sim v \tau_{\rm m} \sim l_{\rm m}$, which is very short and thus potentially compatible with the strong $B$-fields seen immediately downstream in cluster radio relics and supernova thin rims. 

However, exponential growth only takes place for a limited period of time. As the $B$-field grows, magnetic tension starts to play a role, inhibiting vortical motions and field line stretching. This suppression happens when ${\bm B} \cdot \nabla {\bm B} \sim \rho {\bm u} \cdot \nabla {\bm u}$, or when $\langle B^{2} \rangle \sim \rho u_{l_{\rm m}}^{2} \sim (l_{\rm m}/L)^{2/3} \rho \langle u^{2} \rangle \sim {\rm Re}^{-1/2} \rho \langle u^{2} \rangle$, where $u$ is the velocity on the outer scale and we have used ${\rm Re} \sim (L/l_{\rm m})^{4/3}$. Since the inner scale has the least kinetic energy, magnetic fields come into equipartition with it first. However, larger scale motions have more energy, and are not yet quenched; it takes a larger $B$-field to suppress them. As the $B$-field is amplified, ever larger scales are suppressed. If we define a stretching scale $l_{\rm s}(t)$ as the scale at which there is equipartition between solenoidal kinetic energy and magnetic energy, $\rho u_{l_{\rm s}}^{2} \sim \langle B^{2} \rangle(t)$, then \citep{schekochihin07}:
\begin{equation}
\frac{d}{dt} \langle B^{2} \rangle \sim \frac{u_{l_{\rm s}}}{l_{\rm s}} \langle B_{l_{s}}^{2} \rangle \sim \frac{\rho u_{l_{s}}^{3}}{l_{s}} 
\sim \epsilon = {\rm const} \Rightarrow \langle B^{2} \rangle(t) \sim \epsilon t 
\end{equation}
Thus, the magnetic energy quickly transitions from exponential growth to linear growth, with the majority of the amplification taking place in the linear growth phase. During this time, $l_{\rm s} \propto t^{3/2}$, until we reach the outer scale L and achieve equipartition, $\langle B^{2} \rangle \sim \langle u^{2} \rangle$; at this time, the coherence length of the $B$-field (as dictated by magnetic tension) is of order the outer scale. This picture of the turbulent dynamo has been verified in MHD stirring box simulations \citep{schekochihin07,cho09}, and we shall use it in our analysis of $B$-field amplification in the post-shock flow. 

We caution the reader that conditions in the post-shock flow can be quite different and less controlled than in stirring box simulations. For instance, turbulent forcing is anisotropic and impulsive; turbulent forcing does not reach a steady state but decays in the post-shock flow. We have no control over the velocity power spectrum. In our setup, temporal variation maps onto spatial variation; properties of the flow such as turbulence and $B$-fields are a strong function of post-shock distance, and hence highly anisotropic.\footnote{Indeed, we cannot measure the velocity or magnetic power spectrum. It only makes sense to measure power spectra in narrow strips perpendicular to the shock normal, but there is insufficient sampling (particularly at large scales) to produce a statistically significant power spectrum.} Most significantly, RMI induced turbulence does not cascade from large to small scales, but instead is injected impulsively at all scales; it violates locality in k-space. Thus, we might expect the power spectrum of turbulence to be more Burgers-like ($E(k) \propto k^{-2}$) than Kolmogorov-like. While the quantitative details of the small scale dynamo will certainly differ between RMI induced turbulence and that seen in stirring boxes, we expect the broad brush aspects to remain the same. As long as the turbulent energy density is dominated by large scales and vorticity is dominated by small scales (true as long as $E(k) \propto k^{-n}$ with $n>1$), then we expect magnetic energy to grow in an inverse cascade from small to large scales, with magnetic tension mediating a transition between exponential and linear growth. We will now see if these ideas are applicable to our simulation results.

\section{Results} 
\label{sect:results} 

In what follows, we describe the results of 2D simulations with an initially perpendicular field. We will describe the results of varying initial $B$-field geometry and 3D simulations later. We consider $\beta=1,40$. The former is a typical plasma beta for the ISM, and as argued in \S\ref{sect:key_parameters} may even be representative of the outskirts of galaxy clusters, where radio relics are more commonly seen.

The $\beta=40$ case is more representative of the plasma beta in the bulk of the ICM. 

\begin{figure*}
  \begin{subfigure}[b]{0.48\textwidth}
    \centering
    \includegraphics[width=\textwidth]{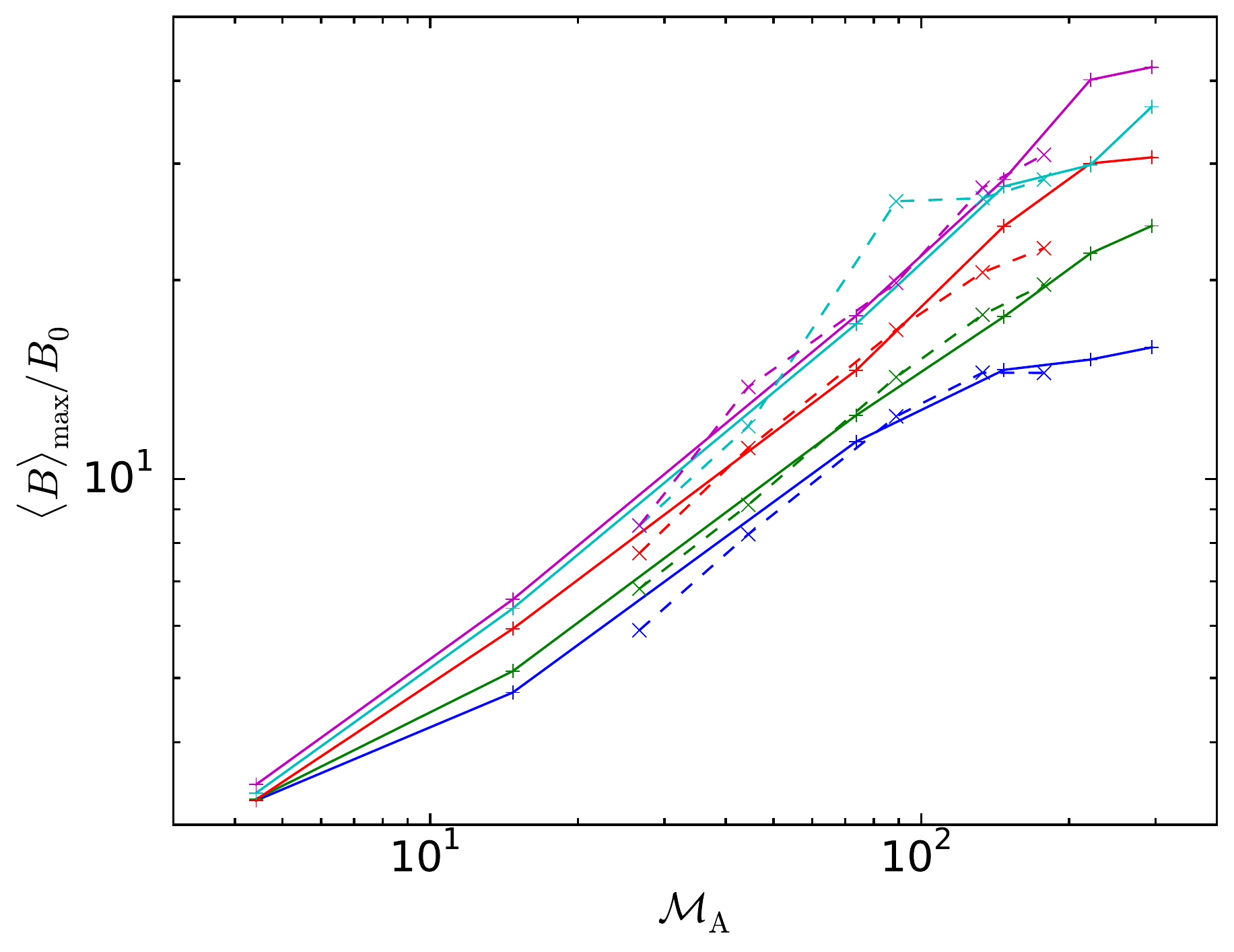}
    \caption{$\langle B\rangle_\mathrm{max}/B_{0}$ vs. $\mathcal{M}_\mathrm{A}$}
    \label{fig:B_amp_vs_macha_aver}
  \end{subfigure}
  \begin{subfigure}[b]{0.48\textwidth}
    \centering
    \includegraphics[width=\textwidth]{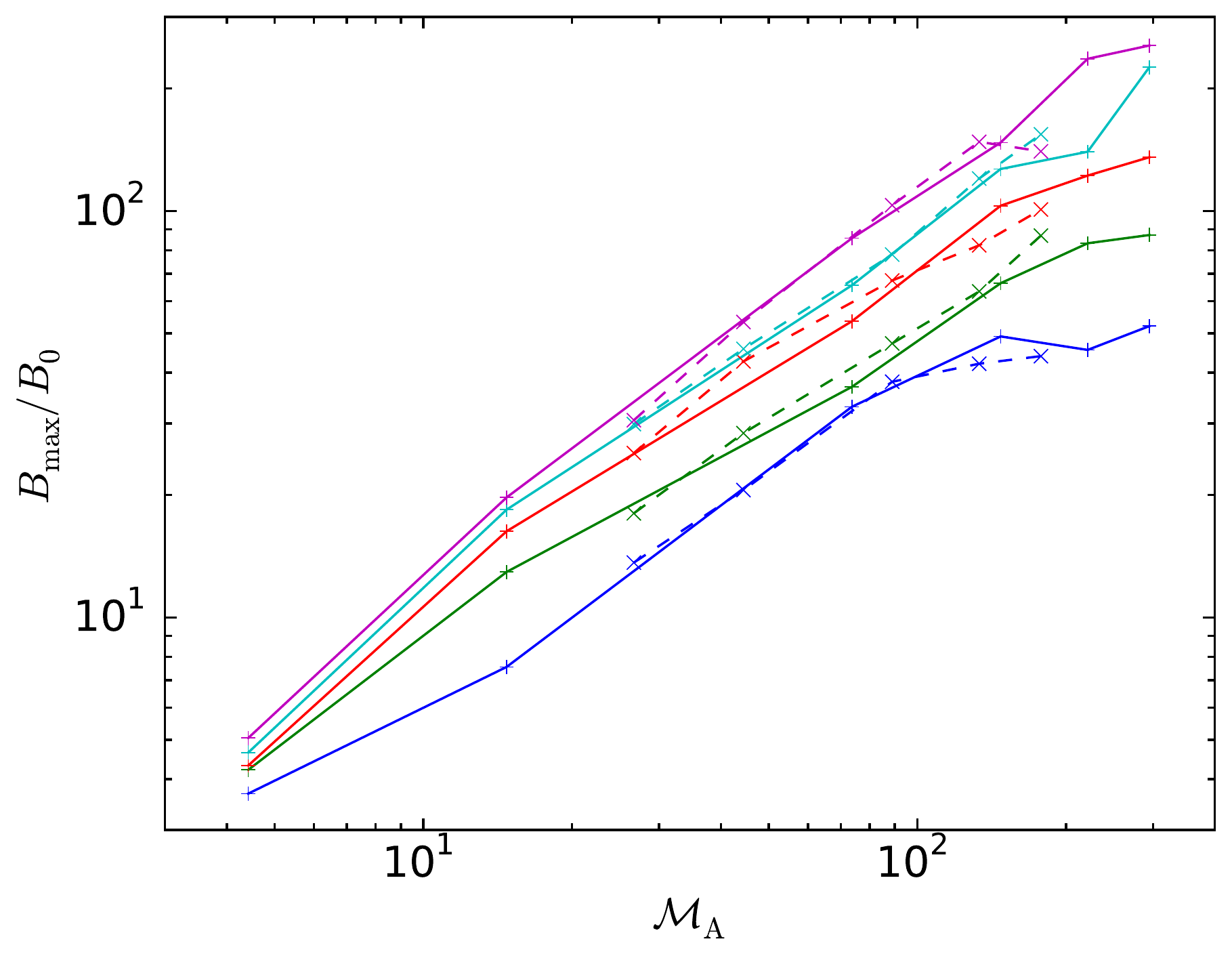}
    \caption{$B_\mathrm{max}/B_{0}$ vs. $\mathcal{M}_\mathrm{A}$}
    \label{fig:B_amp_vs_macha_aver_global}
  \end{subfigure}
  \caption{Scaling of magnetic field amplification with Alfv\'{e}n Mach numbers, for an initially perpendicular field, in 2D. Different colors represent different resolution: blue -- $256\times64$, green -- $512\times128$, red -- $1024\times256$, cyan -- $2048\times512$, magenta -- $4096\times1024$. Solid lines are for plasma $\beta=1$, and dashed lines are for $\beta=40$. The slight flattening seen in the $\beta=40$ case at high $\mathcal{M}_\mathrm{A}$ is due to inadequate box length: higher saturation levels require more time and a longer box. $B_0$ is the initial field, $\langle B\rangle_\mathrm{max}$ refers to the maximum vertically ($y$) averaged field for a strip of distance $x$ from the shock, while $B_\mathrm{max}$ refers to the maximum field in the simulation box.}
  \label{fig:mach_macha_var}
\end{figure*}

\subsection{Dependence on $\mathcal{M}_{\rm A}$} 

Fig. \ref{fig:mach_macha_var} shows how the maximum amplification in a simulation box ($B_{\rm max}/B_0$) and the maximum y-averaged $B$-field $\langle B\rangle_{\rm max} $ (i.e., averaged in strips parallel to the shock; this is thus the maxima of $B$-field profiles such as shown in Fig. \ref{fig:bmag_vs_x_cluster}, \ref{fig:bmag_vs_x_sn}) vary with the Alfv\'en Mach number $\mathcal{M}_{\rm A}$, for a range of resolutions and for $\beta=1,40$ (solid, dashed lines respectively).  We see that amplification increases with resolution, but we eventually reach convergence between the $2048 \times 512$ and $4096 \times 1024$ simulations. We will have more to say about convergence properties later. We see a linear relation $B/B_0 \propto \mathcal{M}_{\rm A}$, as expected \citep{fraschetti13}. This can be easily understood: for a strong shock encountering sufficient density fluctuations (see below) to excite strong turbulence, $U_{\rm B} \sim U_{\rm turb} \propto U_{\rm ram} \sim \rho \mathcal{M}_{\rm A}^{2} v_{\rm A,0}^{2} \sim \mathcal{M}_{\rm A}^{2} U_{\rm B_{0}} \Rightarrow (B/B_0) \propto \mathcal{M}_{\rm A}$. The Alfv\'en Mach number thus determines the amplification factor, while the sonic Mach number $\mathcal{M}_{\rm S}$ determines the absolute value of the peak magnitude field (since $U_{\rm B} \propto U_{\rm turb} \propto U_{\rm ram} \propto \mathcal{M}_{\rm S}^{2}$). 

This also implies that up to factors of order unity, the post-shock field strength is roughly independent of its pre-shock value. As we shall see, once clumping exceeds a threshold value, a significant fraction of the incoming energy flux is converted to turbulence, and the magnetic energy density reaches equipartition with this. The main differences we later find is that for weaker seed fields, the peak field strength is reached further downstream, and field geometry differs when compressional and turbulent amplification dominate. 

\begin{figure}
  \begin{center}
    \includegraphics[width=0.5\textwidth]{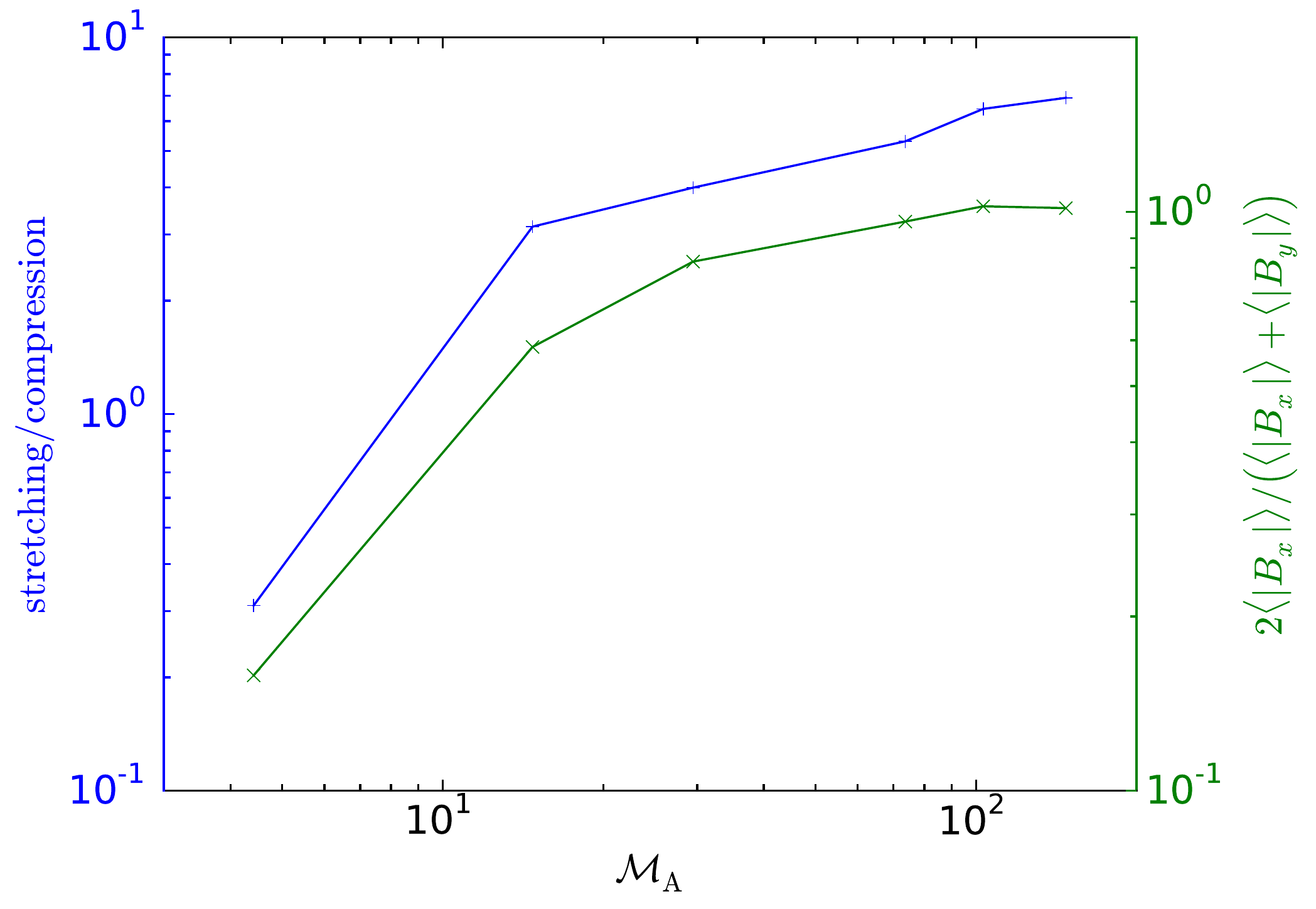}
    \caption{Time averaged ratio of stretching to compression (blue; see equation \ref{eqn:Bfield_amp_terms} for definitions) and magnetic anisotropy (green) vs. $\mathcal{M}_\mathrm{A}$, for the same simulations as in Fig. \ref{fig:mach_macha_var}.}
    \label{fig:stretch_compress_iso_vs_macha}
  \end{center}
\end{figure}

A few other features of these plots are worth noting. $B_{\rm max}/B_0$ is significantly larger than $\langle B \rangle_{\rm max}/B_{0}$, by up to a factor of $\sim 5$, implying strong fluctuations in $B$-field strength. This can also be seen in Figs \ref{fig:2d_slice} and \ref{fig:slice_3d_iso}. These fluctuations are to be expected in the presence of large scale turbulence. Secondly, for $\mathcal{M}_{\rm A} \ge 30$, the $B$-field amplification depends only on $\mathcal{M}_{\rm A}$, and is independent of $\beta$ (note that $\mathcal{M}_{A} \sim 30$ is the lowest Alfv\'en Mach number accessible for $\beta = 40$ and $\mathcal{M}_{\rm S} \ge 4$; we expect the $\beta$ independence to be true down to $\mathcal{M}_{\rm A} \sim 10$). This makes sense: this corresponds to the case where turbulent amplification strongly dominates (Fig. \ref{fig:stretch_compress_iso_vs_macha}). For $\mathcal{M}_{\rm A} \gg 1$, $U_{\rm ram} \gg U_{B_0}, U_{\rm therm}$; the initial $B$-field is dynamically unimportant and $\beta$ scales out of the problem. Note that the curves start to flatten for $\mathcal{M}_{\rm A} > 150$. This is a purely numerical effect: the smaller initial $B$-field means that it takes longer to amplify the fields up to their peak, and our box length is insufficient. Nonetheless, the main trends are clear from this plot. 

\subsection{Compressional vs Turbulent Amplification}

In Fig. \ref{fig:stretch_compress_iso_vs_macha}, we see the time averaged ratio of the stretching $\bm{B} \cdot(\bm{B}\cdot\bm{\nabla})\bm{v}$ and compression $|\bm{B}|^2 \bm{\nabla}\cdot \bm{v}$ terms in equation \ref{eqn:Bfield_amp_terms}, as a function of the Alfv\'en Mach number $\mathcal{M}_{\rm A}$. We see that compression dominates at low $\mathcal{M}_{\rm A}$, while stretching dominates at high $\mathcal{M}_{\rm A}$, with a cross-over at $\mathcal{M}_{\rm A} \sim 10$. Compression has a fixed amplification given by the shock compression ratio $r \sim 2-4$ (depending on the Mach number of the shock) for a perpendicular field, and $[(2 r^{2} +1)/3]^{1/2}$ for an isotropic tangled field. 

This transition from compressional to turbulent amplification has implications for postshock magnetic geometry (green line, Fig. \ref{fig:stretch_compress_iso_vs_macha}). In low $\mathcal{M}_{\rm A}$ shocks, when turbulent motions are suppressed by magnetic tension, the field is relatively static and passively advected across the shock, with the perpendicular field amplified by the shock compression ratio, and the parallel field continuous across the shock, resulting in a field with perpendicular bias (note that our results are for a perpendicular field, which maximizes the contribution of compression; it is smaller for an oblique or tangled field). As turbulent amplification becomes increasingly important, the field becomes increasingly tangled, reaching isotropy at high $\mathcal{M}_{\rm A}$. As we shall see, while the result for field geometry at low $\mathcal{M}_{\rm A}$ is robust, the isotropy seen at high $\mathcal{M}_{\rm A}$ is a 2D artifact. In 3D simulations, far downstream from high $\mathcal{M}_{\rm A}$ shocks, fields have a {\it parallel} bias. 

\begin{figure*}
  \begin{subfigure}[b]{0.48\textwidth}
    \centering
    \includegraphics[width=\textwidth]{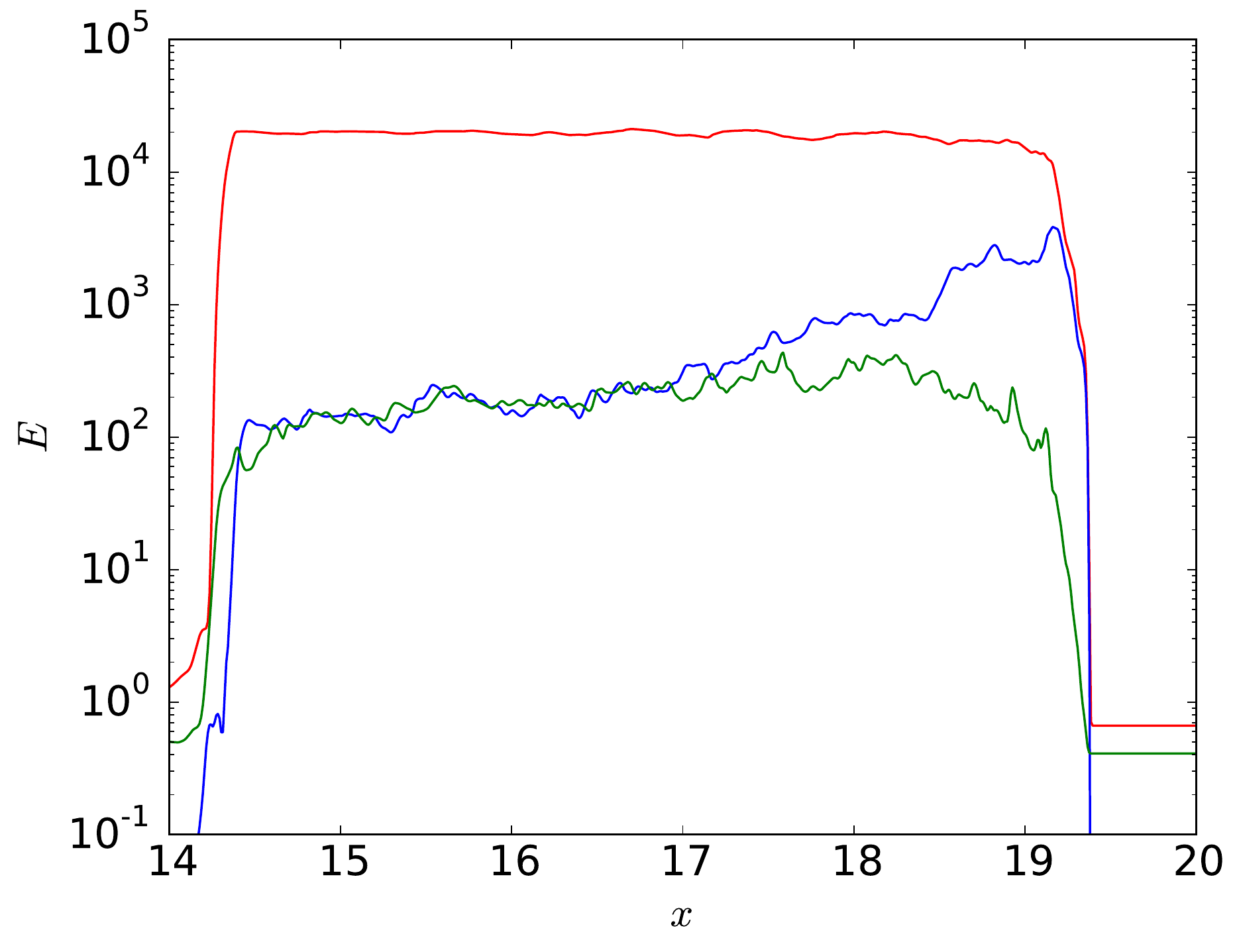}
    \caption{$\mathcal{M}=133, \beta=1$}
    \label{fig:sn_energy_36}
  \end{subfigure}
  \begin{subfigure}[b]{0.48\textwidth}
    \centering
    \includegraphics[width=\textwidth]{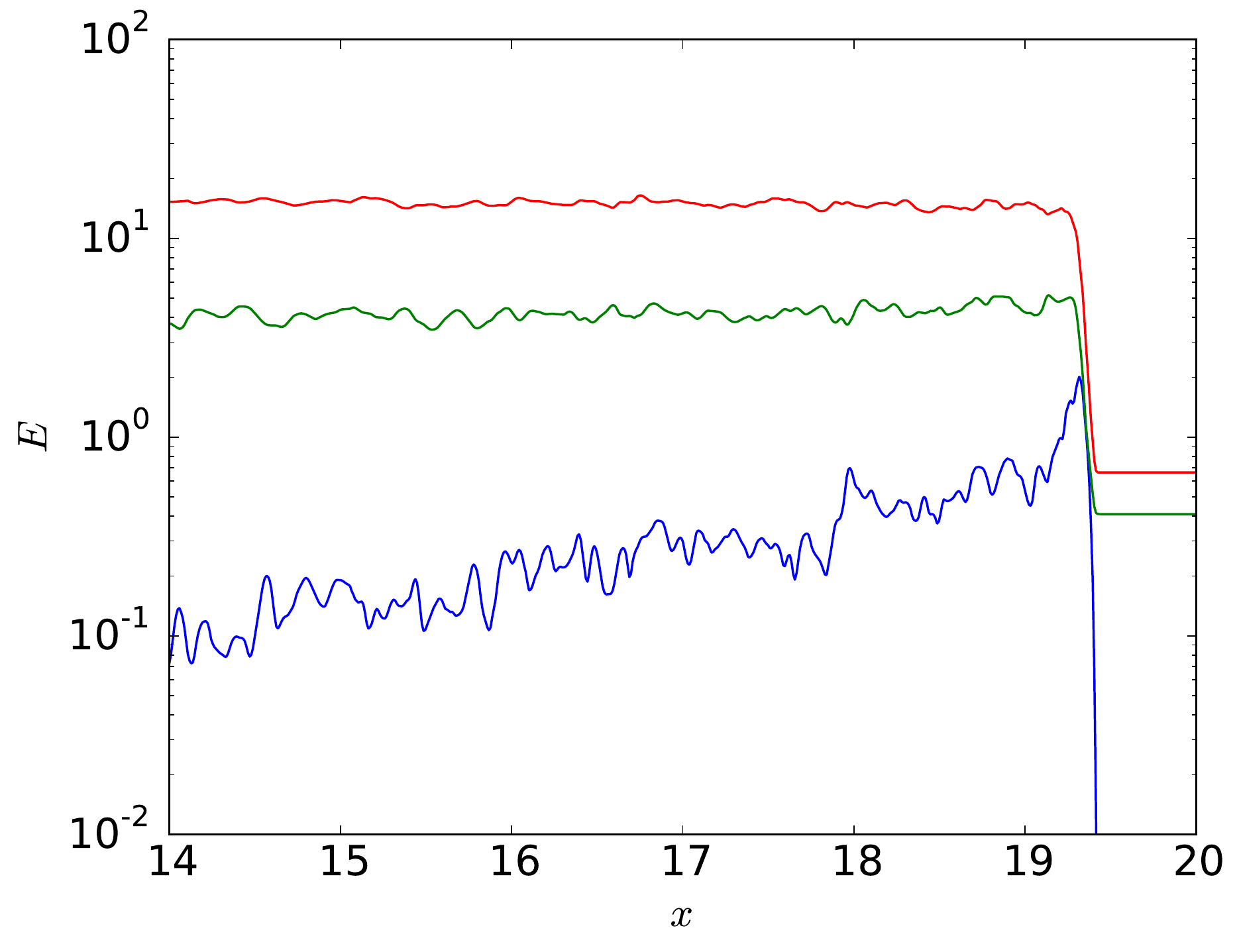}
    \caption{$\mathcal{M}=4, \beta=1$}
    \label{fig:cluster_energy_33}
  \end{subfigure} 
  \caption{Profiles of energy density from 2D simulations: turbulent energy density (blue), magnetic energy density (green), and thermal energy density (red).}
  \label{fig:radial_energy}
\end{figure*}

\subsection{Energy densities}

In Fig. \ref{fig:radial_energy}, we show how the thermal, magnetic and turbulent energy density evolve as a function of distance from the shock. In the high Mach number case, the turbulent energy density\footnote{The turbulent velocity dispersion is $\sigma = v-\bar{v}$. Since the mean downstream velocity $\bar{v}=v_{\rm p}$, this is equivalent to evaluating velocities in the frame of the piston, where the postshock fluid has no net bulk flow.} increases sharply at the shock, slowly decaying due to numerical viscosity and $B$-field growth via the turbulent dynamo (which saps energy from turbulent motions and eventually damps them via magnetic tension). Turbulent and magnetic energy densities reach equipartition. This is consistent with the post-shock lengthscale in Fig. \ref{fig:radial_energy} on which the $B$-field reaches equipartition: 
\begin{equation}
\delta l_{\rm peak} \approx v_{d} (n t_{\rm eddy}) \approx \frac{v_{s}}{4} \left( n \frac{L_{\rm max}}{\sigma} \right) \approx 2.5 n \, L_{\rm max}
\label{eqn:lpeak} 
\end{equation} 
where $n$ is the number of eddy turnovers for the $B$-field to peak, and we have used a downstream bulk velocity of $v_{\rm d}=v_{\rm s}/4$ for a strong shock, a turbulent velocity dispersion $\sigma \sim 0.1 c_{\rm s} \sim 0.1 v_{s}$ (since we see that $U_{\rm turb} \approx 0.01 U_{\rm therm}$ for a strong shock). A value of $n\sim 2$ eddy turnover times is consistent with Fig. \ref{fig:radial_energy}(a), (where $L_{\rm max} =0.5$), though note that $n$ increases in cases where the initial seed field is weaker, since a longer time during the linear growth phase is required. 
The turbulent and magnetic energy densities then slowly decline together in tandem due to a combination of numerical viscosity and resistivity. By contrast, for the low Mach number case, turbulence is energetically subdominant, and the turbulent dynamo is never important. Instead, the pre-existing magnetic field is simply compressionally amplified at the shock (we have verified this directly by comparing the compressional and stretching terms).The low Mach number, rather than magnetic tension is the reason for the low levels of turbulence: a $\mathcal{M}=4$ shock with an initially weak field ($\beta \approx 50$) shows comparably low levels of turbulence. 

\begin{figure*}
  \begin{subfigure}[b]{0.475\textwidth}
    \centering
    \includegraphics[width=\textwidth]{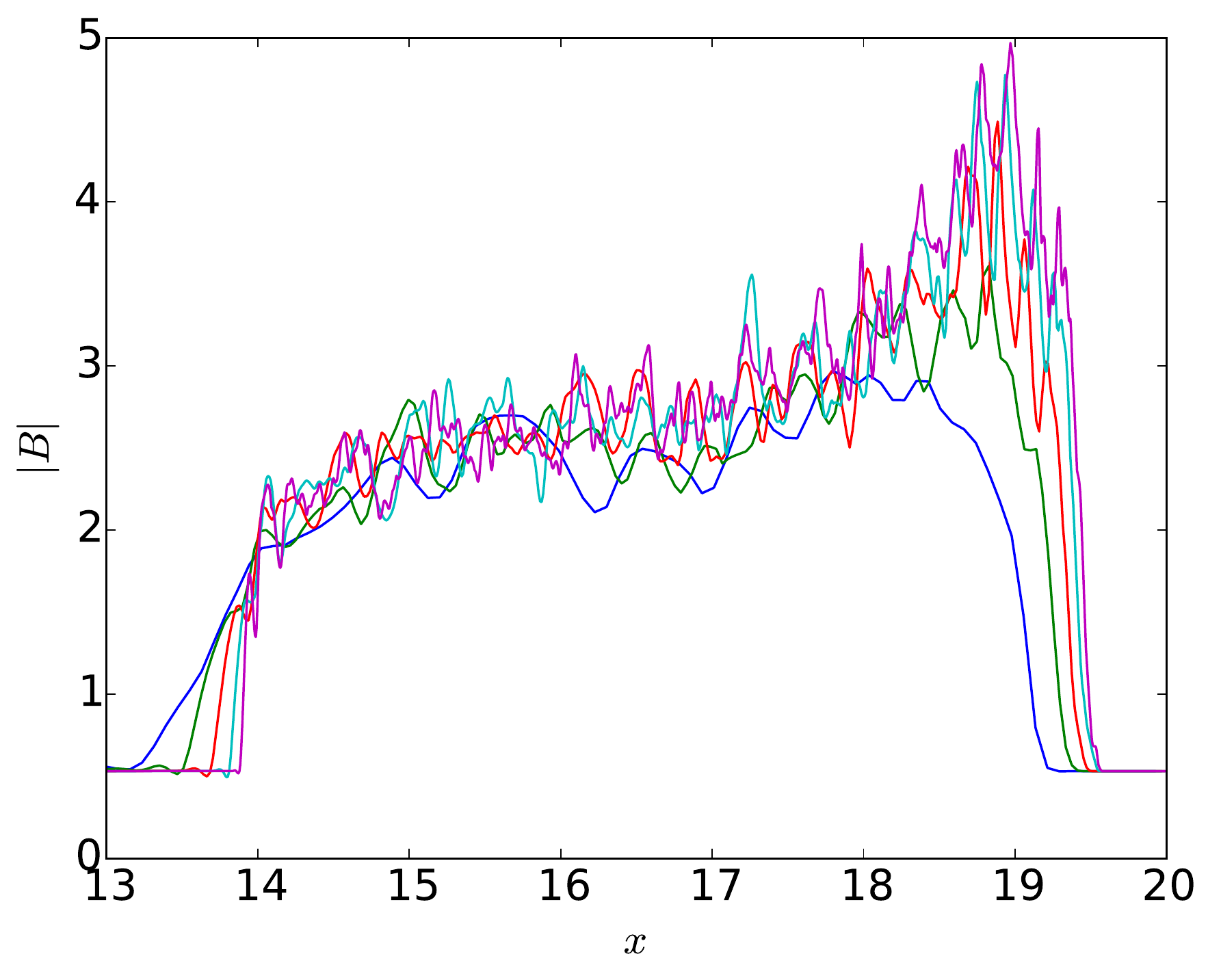}
    \caption{Magnetic field strength, $\mathcal{M}=4, \beta=40$}
    \label{fig:bmag_vs_x_cluster}
  \end{subfigure}
  \begin{subfigure}[b]{0.485\textwidth}
    \centering
    \includegraphics[width=\textwidth]{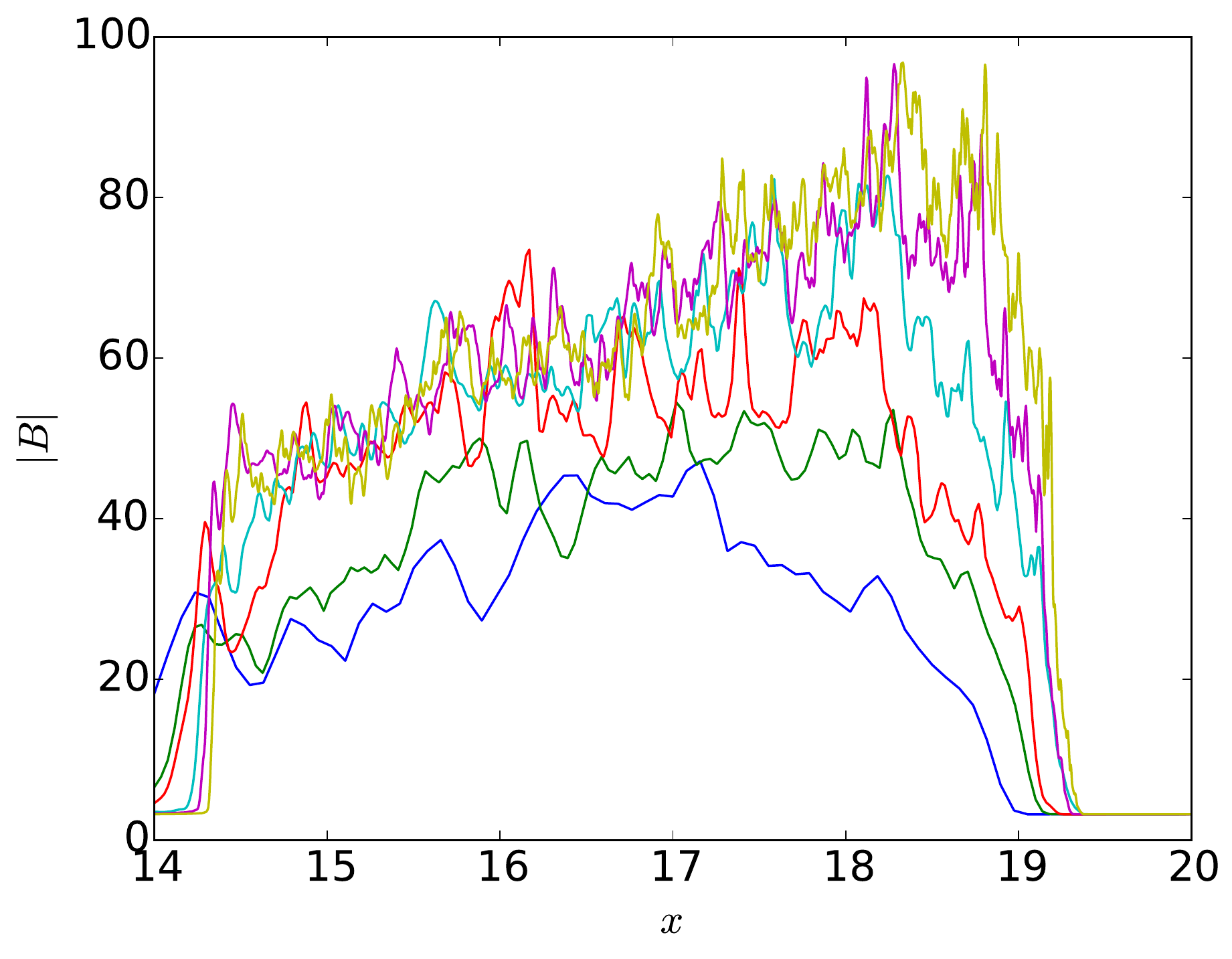}
    \caption{Magnetic field strength, $\mathcal{M}=133, \beta=1$}
    \label{fig:bmag_vs_x_sn}
  \end{subfigure}
  \\
  \begin{subfigure}[b]{0.48\textwidth}
    \centering
    \includegraphics[width=\textwidth]{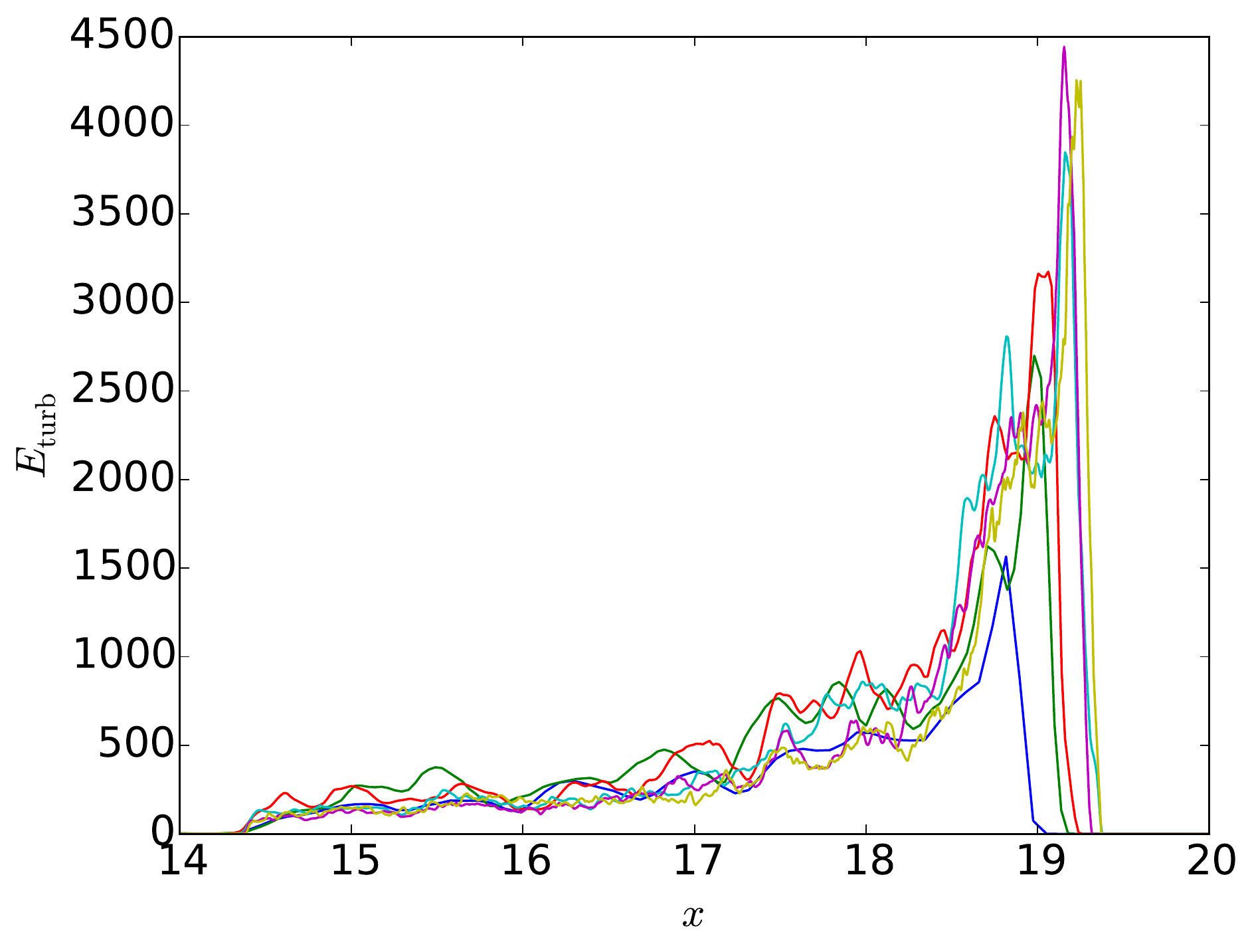}
    \caption{Turbulent energy density, $\mathcal{M}=133, \beta=1$}
    \label{fig:turb_vs_x_sn}
  \end{subfigure}
  \begin{subfigure}[b]{0.48\textwidth}
    \centering
    \includegraphics[width=\textwidth]{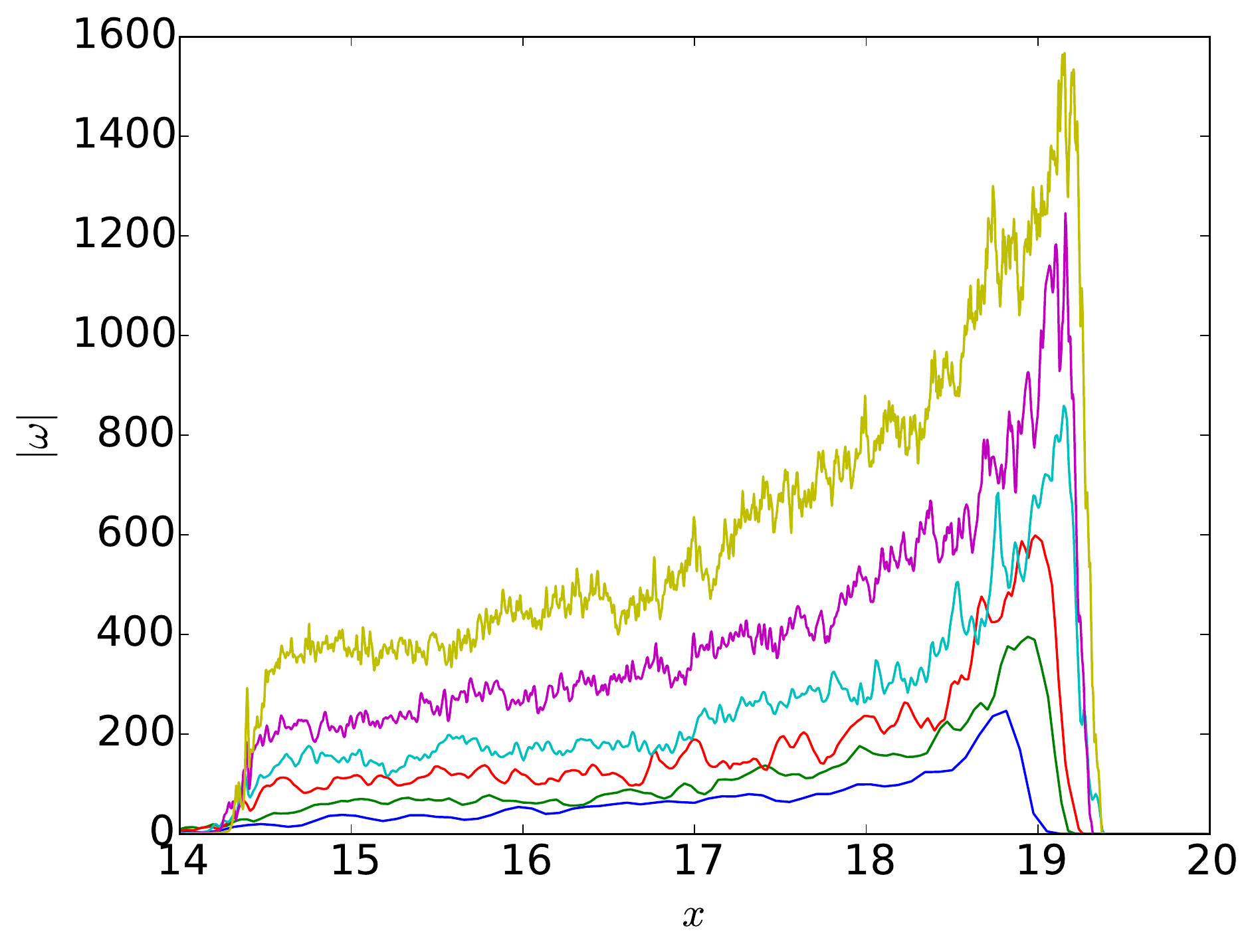}
    \caption{Vorticity magnitude, $\mathcal{M}=133, \beta=1$}
    \label{fig:vort_vs_x_sn}
  \end{subfigure}
  \caption{Profiles of magnetic field strength (a, b), turbulent energy density (c), and vorticity (d), where different colors represent different resolutions: blue -- $256\times64$, green -- $512\times128$, red -- $1024\times256$, cyan -- $2048\times512$, magenta -- $4096\times1024$, yellow -- $8192\times2014$. The simulations are converged in magnetic and turbulent energy density, but not in vorticity.}
  \label{fig:bmag_vort_profile}
\end{figure*}

\subsection{$B$-field and vorticity profiles; convergence}

Fig. \ref{fig:bmag_vort_profile} shows profiles of magnetic field strength, turbulent energy density and vorticity as a function of post-shock distance, for different resolution runs. Naively, one might expect $B$-fields and vorticity to have broadly similar profiles, since the vorticity equation for $\bm{\omega}$ and the induction equation for $\bm{B}$ are mathematically very similar\footnote{Note, however, that since $\bm{\omega} = \nabla \times \bm{u}$, the vorticity equation has terms $\nabla \times (\bm{\omega} \times \bm{u})$ which are non-linear in velocity, which is not true for the induction equation for $\bm{B}$.}. In fact, they look extremely different. As we have seen before, the $B$-field rises rapidly after the shock and subsequently slowly decays due to numerical resistivity. Fig. \ref{fig:bmag_vort_profile} shows that magnetic field profiles are reasonably well converged, up to $\mathcal{M}_{\rm A} \approx 130$ (higher $\mathcal{M}_{\rm A}$ implies more turbulent amplification and more stringent resolution requirements), as one might have expected from Fig. \ref{fig:mach_macha_var}. Similarly, the profiles of turbulent energy density are numerically converged -- as expected, since $U_{\rm turb}$ and $U_{\rm B}$ reach equipartition. Besides the vertically averaged profiles, we have also examined the probability density function (PDF) of $B$-fields in the simulation boxes (not shown), and found that they are converged. Note that in this case, our fiducial resolution of $2048 \times 512$ corresponds to $\sim 50$ cells over the coherence length $L$. Besides increasing resolution, the Reynolds number may be formally increased by increasing $L_{\rm max}$ at fixed resolution. We have performed calculations where we increase $L_{\rm max}$ by a factor of $5$, and find similar maximum $B$-fields. 

However, Fig. \ref{fig:bmag_vort_profile} shows that vorticity profiles are very different from $B$-field profiles, and moreover are {\it not} numerically converged. Vorticity rises sharply at the shock, and subsequently decays due to numerical viscosity, since there is no further driving or source of free energy in the post-shock fluid (unlike the case for $B$-fields, which can feed off the turbulent dynamo to keep growing). Moreover, the lack of convergence shows that vorticity is essentially unresolved and dominated by grid scales; it grows continuously with increasing resolution. Vorticity would be converged if we implemented a  larger viscous term in our equations; to maximize our available dynamic range, we have not chosen to do so. Fortunately, this lack of convergence does not affect the main quantity that we are interested in, the magnetic field $\langle {\bm B}(x) \rangle$. It {\it does} affect the rate of initial exponential $B$-field growth, which one expects to have an e-folding time $\tau \sim \omega^{-1}$. We have directly verified this expectation for growth time in the exponential phase, particularly for the high $\mathcal{M}_{\rm A}$ shocks in Fig. \ref{fig:mach_macha_var}, when the exponential phase is most extended. One can see these different rates of post-shock $B$-field growth for different resolutions in Fig. \ref{fig:bmag_vort_profile}. However, the saturated, maximum $B$-field amplitude is determined by equipartition with turbulence, whose energy density is dominated by gas motions at large scales, rather than grid scale dynamics. Resolving small scale motions (which are quickly stabilized by magnetic tension) is not important. 

\begin{figure*}
  \begin{subfigure}[b]{0.435\textwidth}
    \centering
    \includegraphics[width=\textwidth]{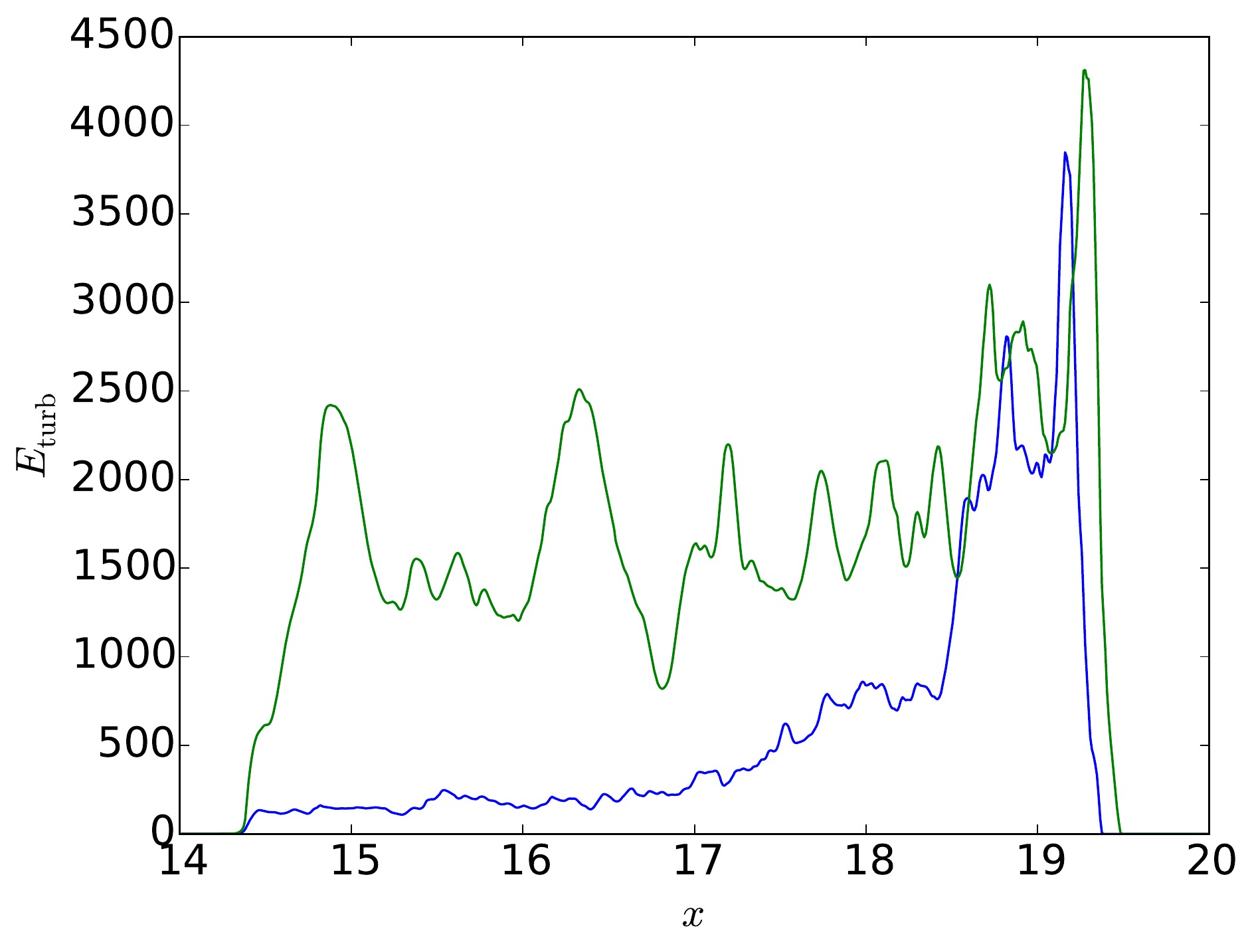}
    \caption{$E_\mathrm{turb}$ in cases of hydro vs. MHD}
    \label{fig:turbulent_energy_hydro_mhd}
  \end{subfigure}
  \begin{subfigure}[b]{0.465\textwidth}
    \centering
    \includegraphics[width=\textwidth]{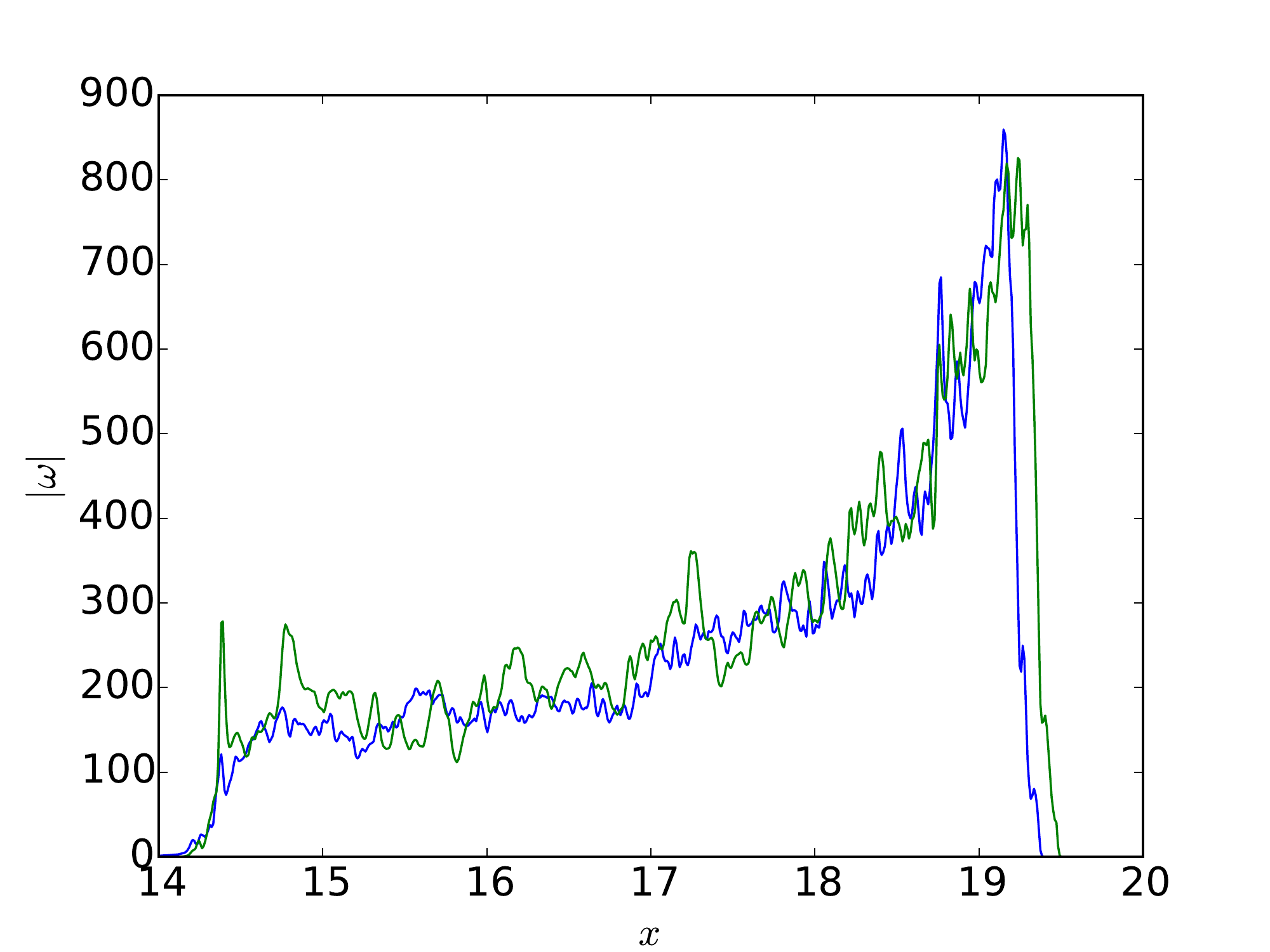}
    \caption{$|\omega|$ in cases of hydro vs. MHD}
    \label{fig:vort_hydro_mhd}
  \end{subfigure}
  \caption{Left panel: comparison of turbulent energy densities for a ${\mathcal M}_{\rm S}=133, \beta=1$ MHD (blue) and hydro (green) simulations. The turbulent energy density decays significantly in the MHD case due to suppression by magnetic tension. Right panel: comparison of vorticity profiles for the same set of MHD and hydro simulations. By contrast, these are very similar, since vortical decay is controlled by numerical dissipation.}
  \label{fig:MHD_vs_hydro}
\end{figure*}

Many papers on RMI $B$-field amplification at shocks in the literature do not present convergence studies. However, by the criteria we present, we believe most studies on planar shocks should have converged maximum $B$-fields. An interesting exception, where a convergence study was performed and results were {\it not} converged, was presented by \citet{guo12}. They performed high-resolution 2D simulations of supernova blast waves propagating through a medium with $L_{\rm max}=3$ pc, and a resolution of $7.5 \times 10^{-3} \mathrm{pc}$; since $L_{\rm max}/\Delta x = 400$, they should be amply converged by our criterion. However, when the forward shock has a radius of $\sim 5$ pc, they analyzed a rim within $\sim 0.3 \rm{pc}$ of the shock front and found that the total magnetic energy increased continuously with resolution. The $B$-field PDF in this region was also not converged. This is to be expected: the narrow region behind the shock is still in the exponential growth phase; convergence in this region requires much higher resolution, since the growth rate is proportional to the (unconverged) vorticity behind the shock. We also note that since $r_{\rm shock} \sim L_{\rm max}$, the peak downstream $B$-field may not be captured (cf equation \ref{eqn:lpeak}); also, curvature of the shock front may become important. 

It is also interesting to compare how gas motions differ in hydrodynamic and MHD simulations (Fig. \ref{fig:MHD_vs_hydro}). One might expect magnetic tension to inhibit small scale motions, and thus for vorticity profiles in MHD and hydro simulations to differ significantly. In fact, they are broadly similar. This is consistent with vortical suppression and decay being chiefly controlled by numerical dissipation. By contrast, the turbulent energy density profiles are very different in hydrodynamic and MHD simulations. The hydrodynamic simulations show that the turbulent energy density remains strong (with large scale fluctuations) after the shock. By contrast, the MHD simulations for the same sonic Mach number show a turbulent energy density which continuously declines to lower levels. The turbulence transfers energy to the magnetic field, and magnetic tension starts to stabilize increasingly larger scales as the $B$-fields approach equipartition. 

\begin{figure*}
  \begin{subfigure}[b]{0.45\textwidth}
    \centering
    \includegraphics[width=\textwidth]{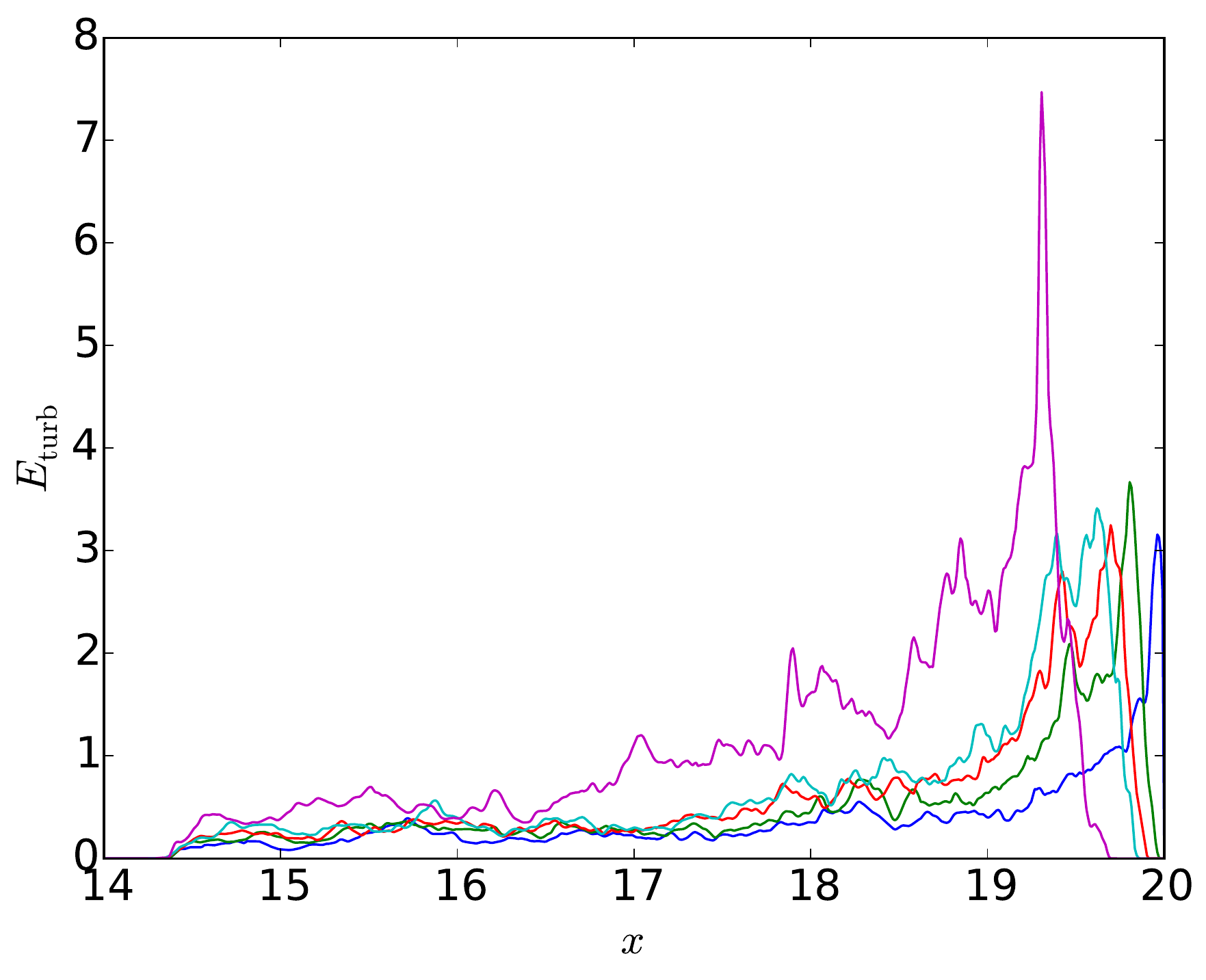}
    \caption{Turbulent energy, $\mathcal{M}=4, \beta=40$}
    \label{fig:eturb_mach3_cluster_clump}
  \end{subfigure}
  \begin{subfigure}[b]{0.45\textwidth}
    \centering
    \includegraphics[width=\textwidth]{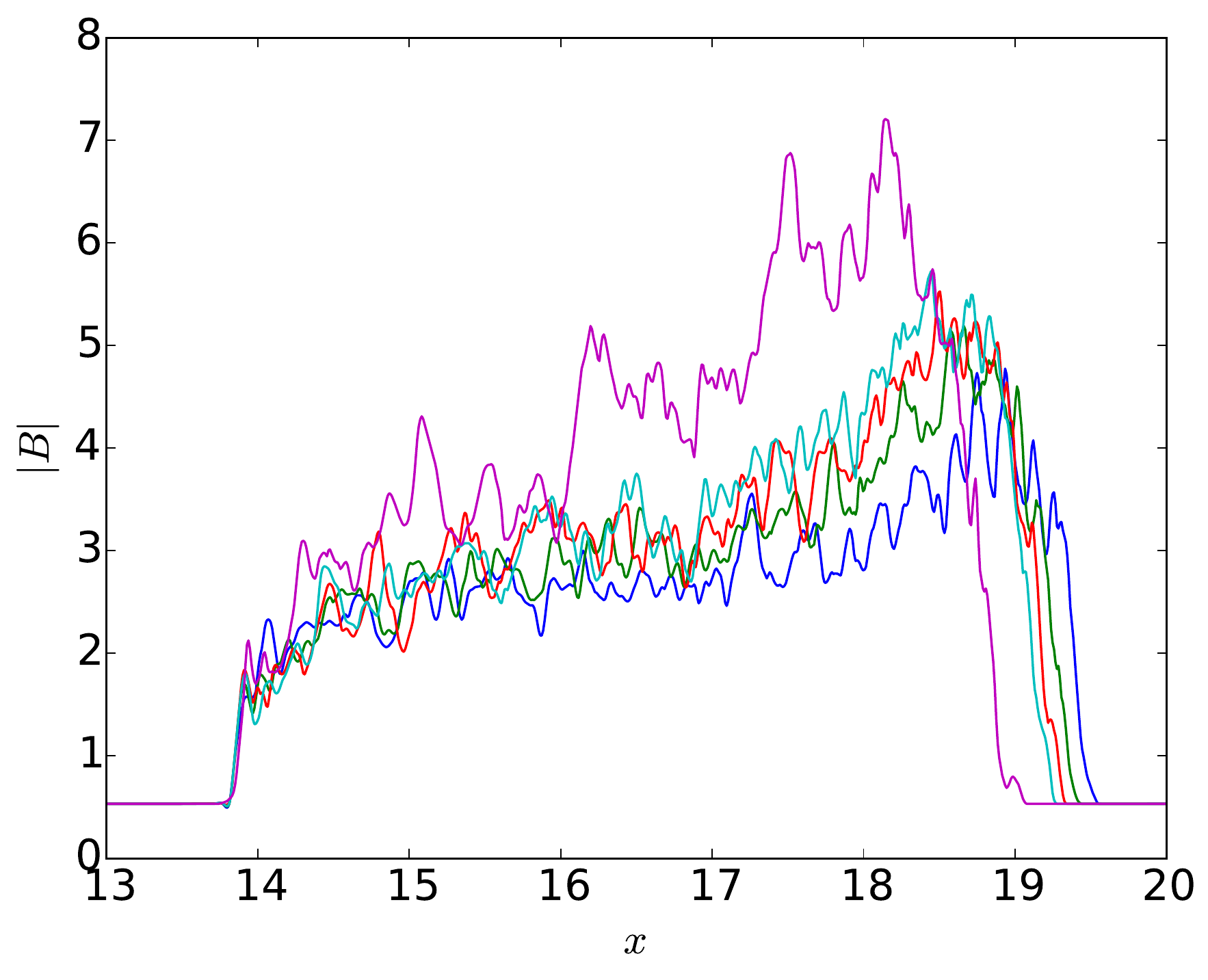}
    \caption{Magnetic field strength, $\mathcal{M}=4, \beta=40$}
    \label{fig:bmag_mach3_cluster_clump}
  \end{subfigure}
  \\
  \begin{subfigure}[b]{0.455\textwidth}
    \centering
    \includegraphics[width=\textwidth]{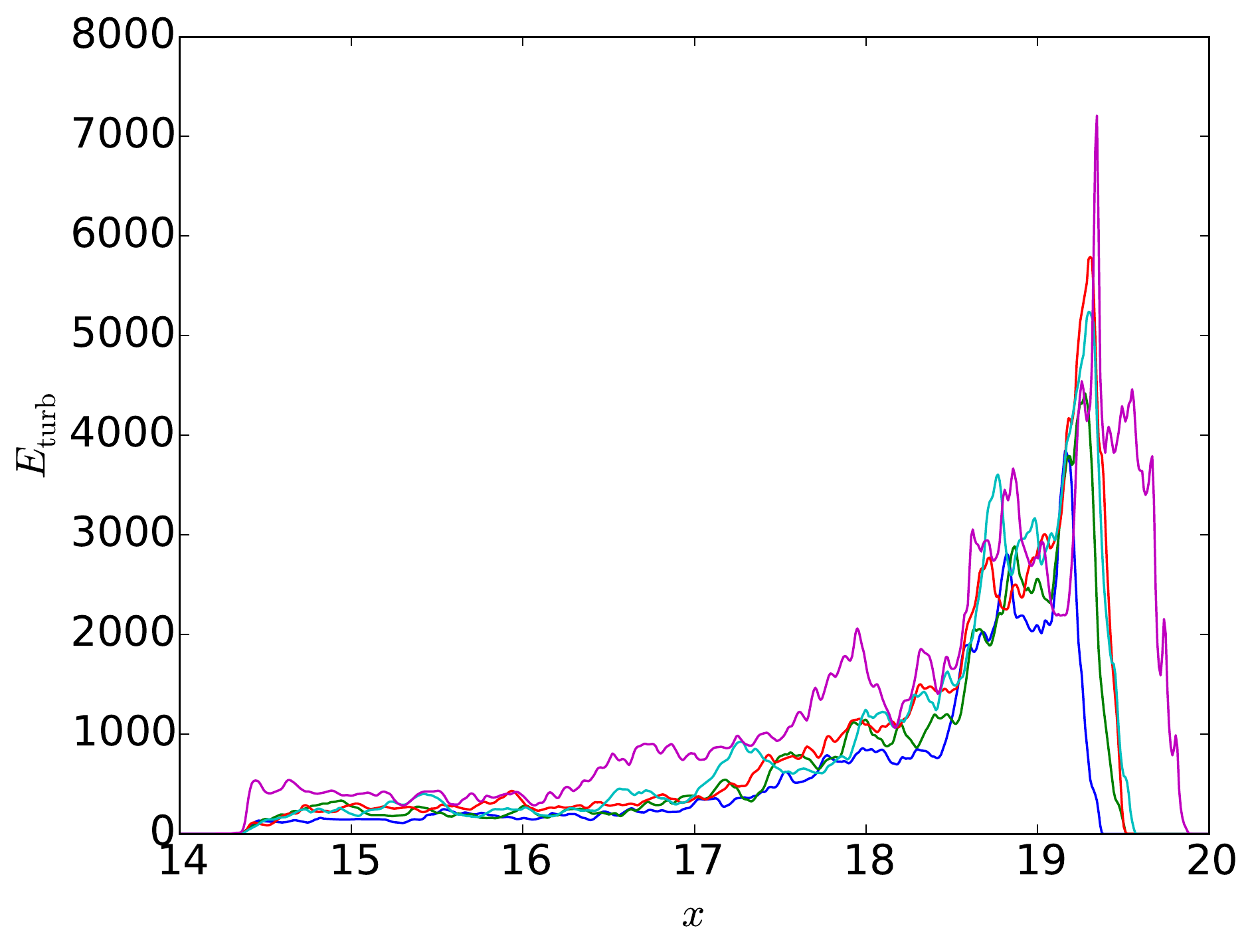}
    \caption{Turbulent energy, $\mathcal{M}=133, \beta=1$}
    \label{fig:eturb_mach100_sn_clump}
  \end{subfigure}
  \begin{subfigure}[b]{0.445\textwidth}
    \centering 
    \includegraphics[width=\textwidth]{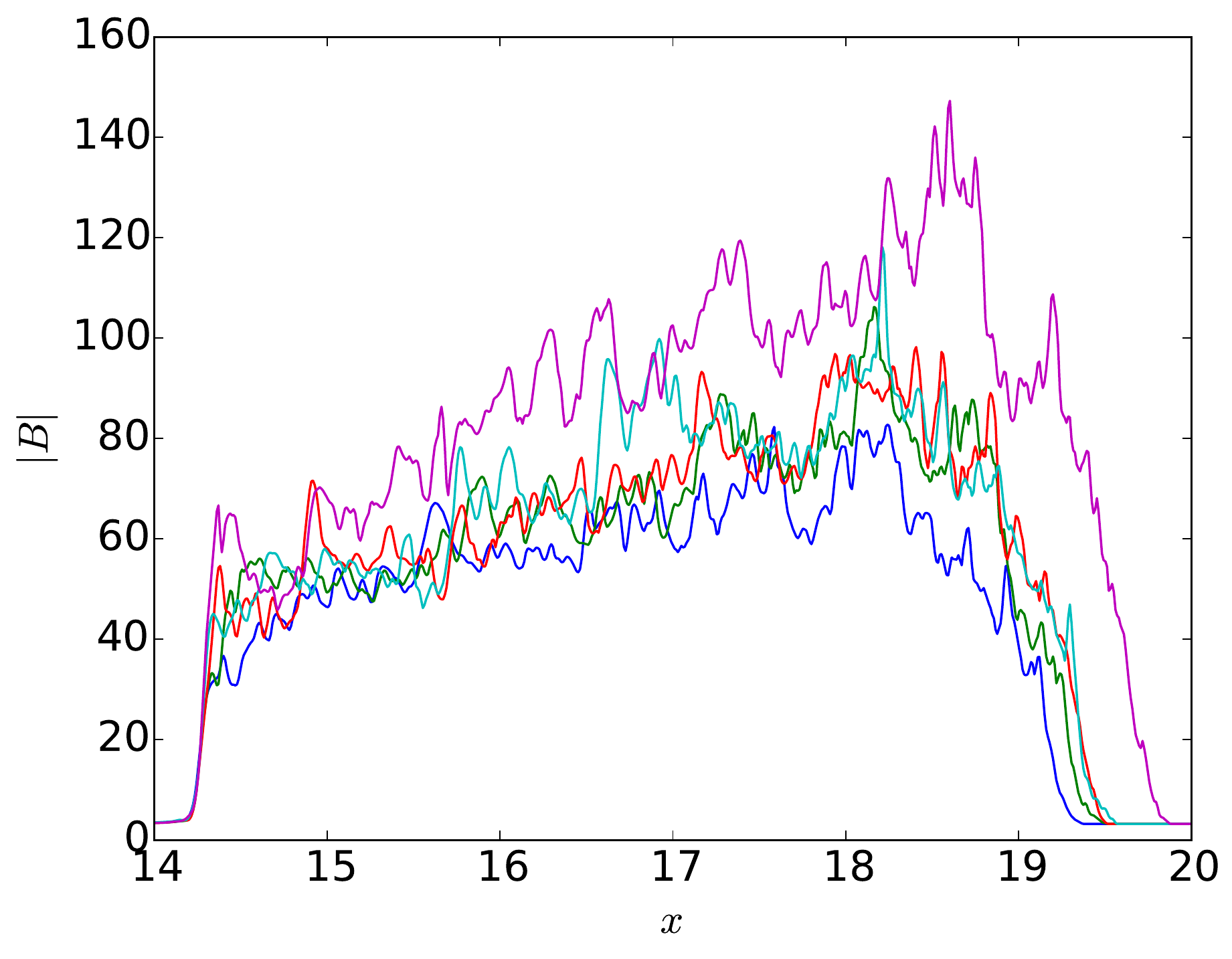}
    \caption{Magnetic field strength, $\mathcal{M}=133, \beta=1$}
    \label{fig:bmag_mach100_sn_clump}
  \end{subfigure}
  \caption{Dependence of turbulent energy and magnetic field strength on gas clumping factor $C = \langle\rho^2\rangle/\langle\rho\rangle^2$: blue -- $C=1.5$, green -- $C=2.0$, red -- $C=2.5$, cyan -- $C=3.0$, magenta -- $C=10$. Above a threshold value of $C \approx 1.5$, these quantities show only a mild dependence on the clumping factor.}
  \label{fig:clump}
\end{figure*}

\subsection{Impact of density inhomogeneities}

In Fig. \ref{fig:clump}, we show the dependence of the turbulent energy and $B$-field profiles on the gas clumping factor $C = \langle \rho^{2} \rangle/ \langle \rho \rangle^{2}$. Characteristic values one might expect for the ISM \citep{ridge06,wong08} and the ICM \citep{simionescu11,zhuravleva13} are $C \sim 2$, albeit with potentially significant scatter and spatial variation. Fortunately, our calculations show that over a broad range of clumping factor $C = 1.5-10$, there is only mild variation of vorticity and $B$-field amplitudes, both for strong and weak shocks. Thus, our results are relatively insensitive to assumptions about gas clumping, which is fortunate, since (particularly in the ICM outskirts) this is a poorly constrained quantity. This saturation of baroclinic vorticity generation and consequently of the turbulent dynamo at relatively small clumping factors was already seen by \citet{inoue13}. They noted this saturation is consistent with expectations from the simple linear analysis by \citet{richtmyer60}, where the turbulent velocity dispersion is:
\begin{equation}
\Delta v \approx \frac{ \Delta \rho/\rho}{(1+ \Delta \rho/\rho)} \langle v_{\rm sh} \rangle (1- \eta) 
\end{equation} 
where $\langle v_{\rm sh} \rangle$ is the shock speed and $\eta$ is the ellipticity of density fluctuations (which vanishes for the isotropic conditions we have assumed). Once rms density fluctuations are of order unity, this expression implies asymptotic turbulent velocities $\Delta v \sim v_{\rm sh} \sim c_{s,2}$, i.e. postshock turbulent energy densities which are comparable (to within a factor of a few) of thermal energy densities. While this is true of hydrodynamic simulations, and MHD simulations in the immediate postshock region, we have already seen that MHD simulations show considerable decline in $U_{\rm turb}$ due to transfer of energy to $B$-fields and subsequent magnetic tension. Thus, when $B$-fields reach equipartition with turbulence, turbulent velocities are much smaller (in our simulations with ${\mathcal M}_{\rm A} = 133$, we see $\sigma \sim 0.1 c_{s}$, and $U_{\rm turb} \sim 10^{-2} U_{\rm therm}$).

\begin{figure*}
  \begin{subfigure}[b]{0.45\textwidth}
    \centering
    \includegraphics[width=\textwidth]{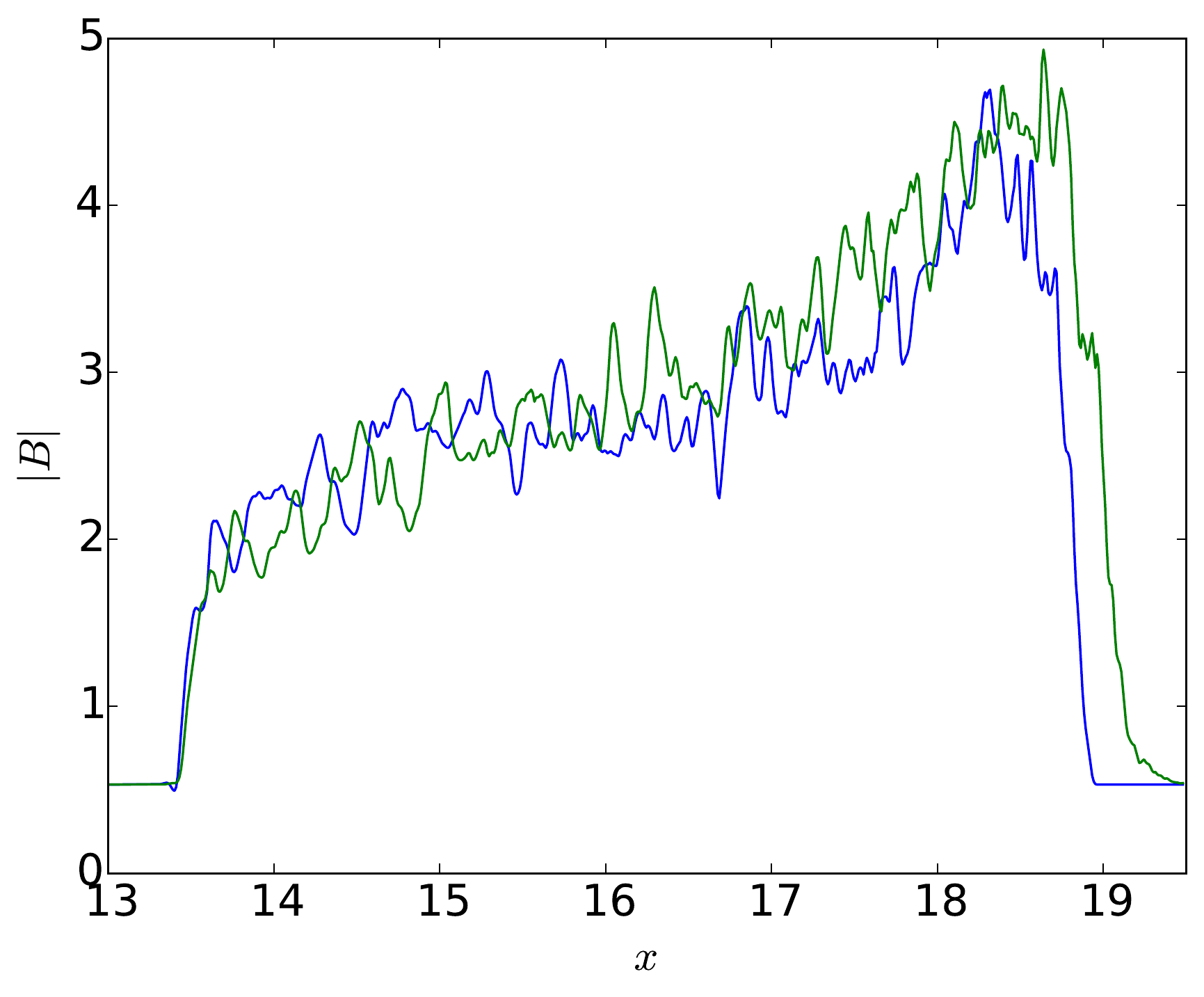}
    \caption{Magnetic field strength, $\mathcal{M}=4, \beta=40$}
    \label{fig:bmag_cluster_initvel}
  \end{subfigure}
  \begin{subfigure}[b]{0.45\textwidth}
    \centering
    \includegraphics[width=\textwidth]{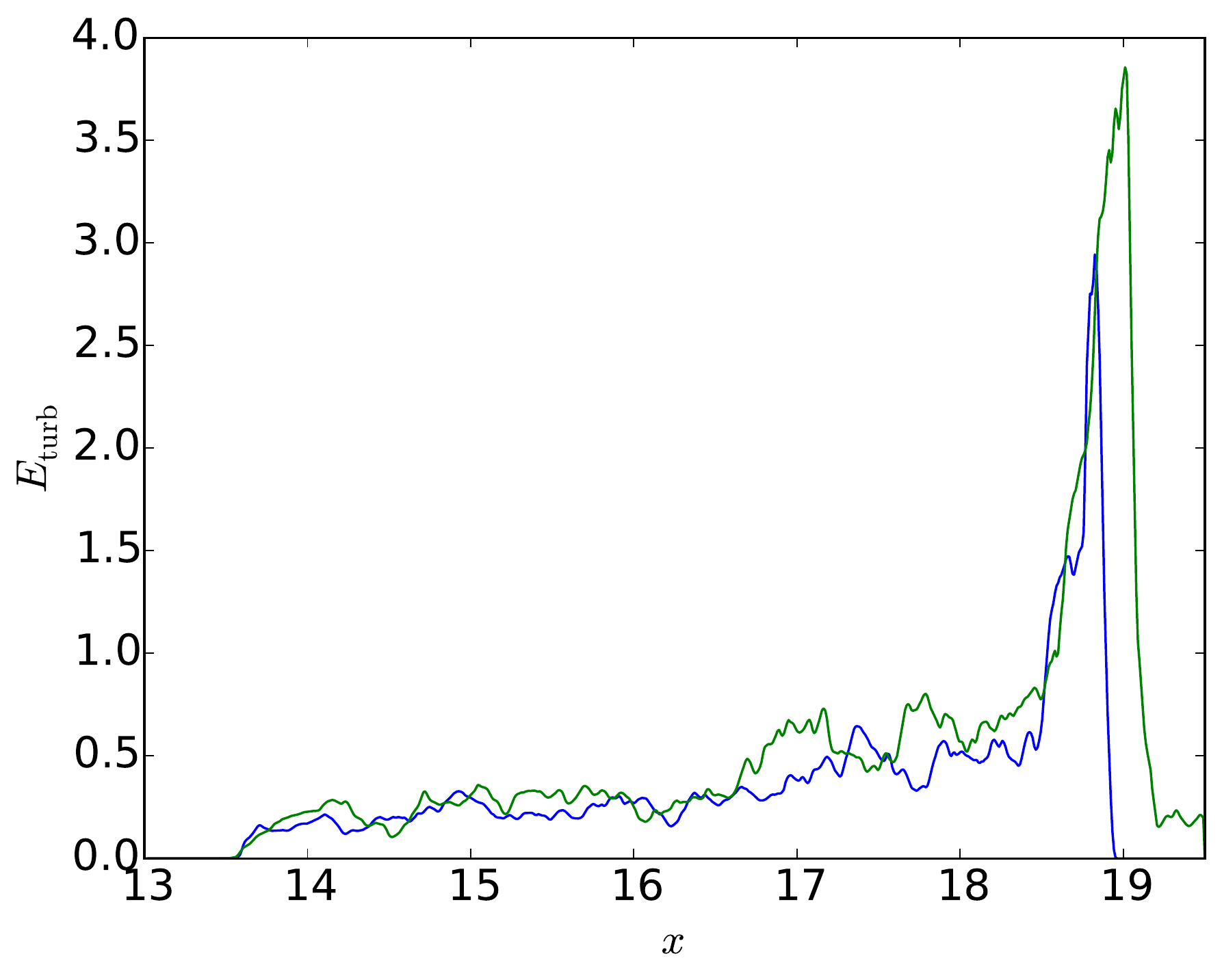}
    \caption{Turbulent energy, $\mathcal{M}=4, \beta=40$}
    \label{fig:eturb_cluster_initvel}
  \end{subfigure}
  \\
  \begin{subfigure}[b]{0.45\textwidth}
    \centering
    \includegraphics[width=\textwidth]{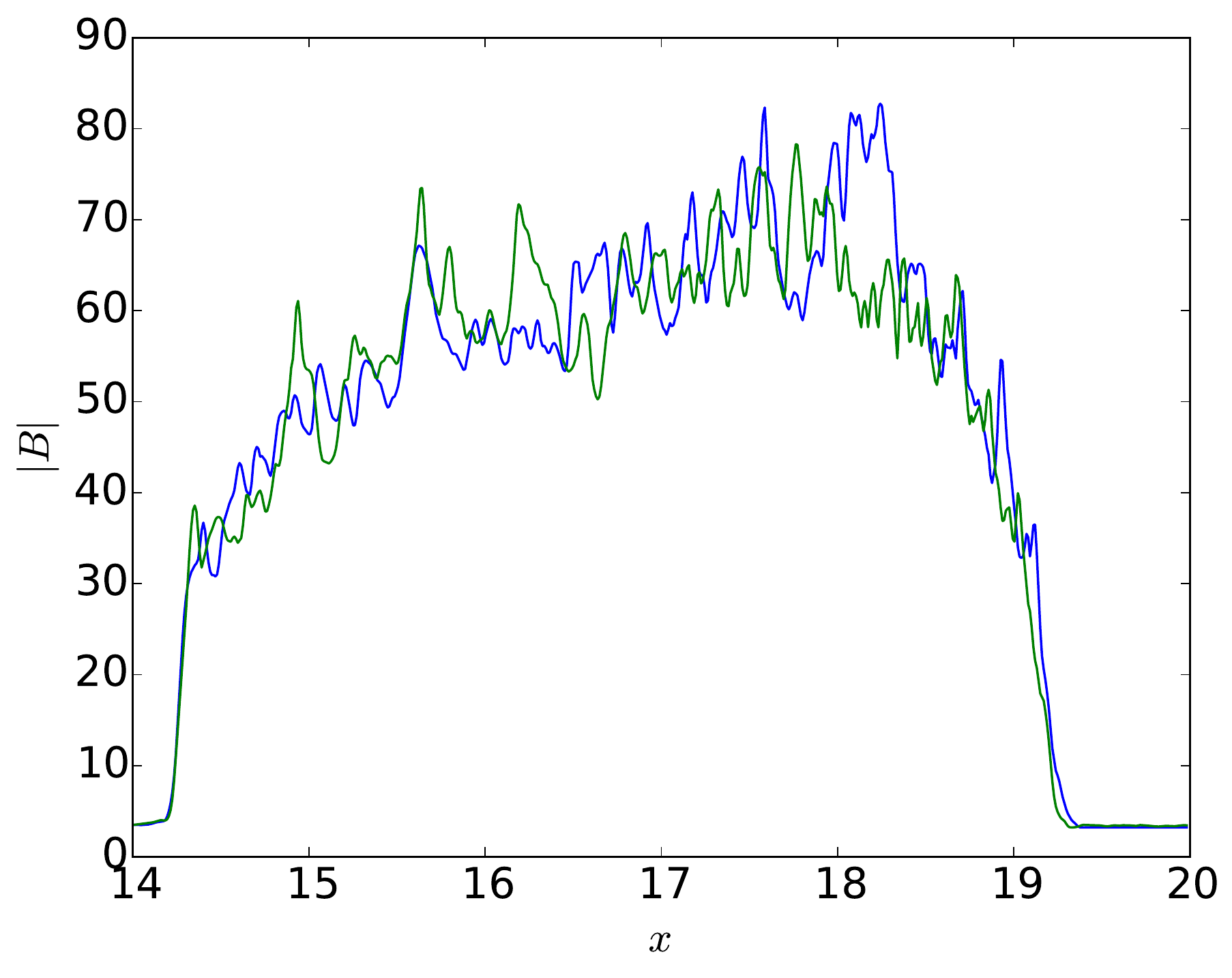}
    \caption{Magnetic field strength, $\mathcal{M}=133, \beta=1$}
    \label{fig:bmag_sn_initvel}
  \end{subfigure}
  \begin{subfigure}[b]{0.45\textwidth}
    \centering
    \includegraphics[width=\textwidth]{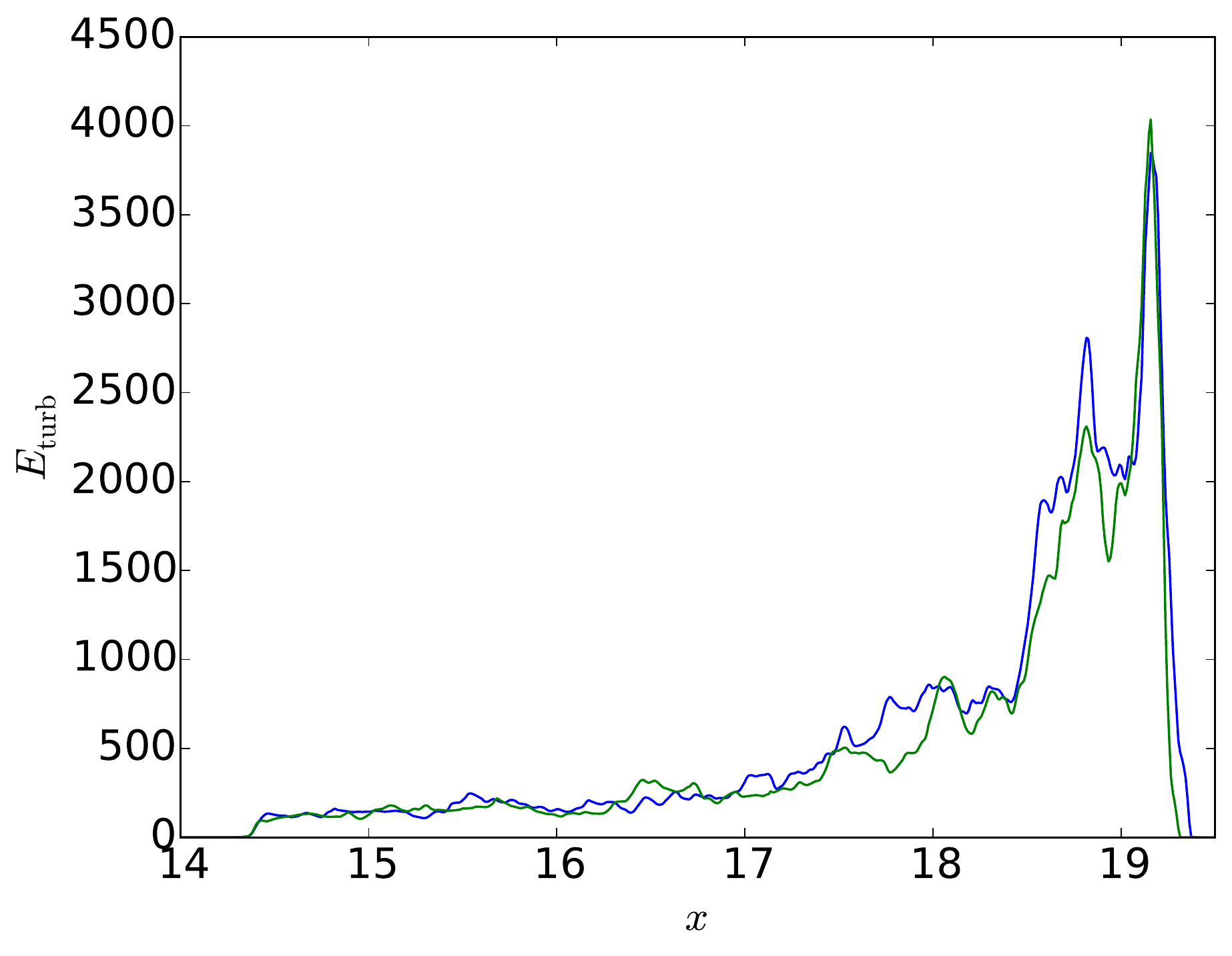}
    \caption{Turbulent energy, $\mathcal{M}=133, \beta=1$}
    \label{fig:eturb_sn_initvel}
  \end{subfigure}
  \caption{Profiles of magnetic field strength and vorticity magnitude, where the green lines have pre-existing solenoidal turbulent velocities of $\mathcal{M}_{\rm S} = 0.5$, and the blue lines indicate an initially static field.}
  \label{fig:bmag_vort_initvel}
\end{figure*}

\subsection{Initial velocity field}

Thus far, we have ignored any initial velocity fluctuations in the pre-shock medium, arguing that turbulent fluctuations in the post-shock fluid are much larger. We now check this explicitly by considering a box where the pre-shock medium has a subsonic (${\mathcal{M}_{\rm S}} \sim 0.5$), solenoidal (and thus incompressible) velocity field. The Mach number of turbulent velocities is what can be expected in the ISM/ICM, to within a factor of a few. To avoid perturbing the imposed density field, we only initialize the velocities a short distance ahead of the shock front as it propagates across the box (note that shock crossing times can be comparable to eddy turnover times, since $L_{\rm box}/L_{\rm max} = 40$). The differences between the static and dynamic initial conditions are shown in Fig. \ref{fig:bmag_vort_initvel}; in all cases, the differences are relatively minor. This reinforces our point that baroclinically generated vorticity is more important than amplification of pre-existing vorticity. 

Most of the simulations in this paper involve perpendicular (or parallel) $B$-fields, which are force free. The exception is the isotropic $B$-fields in 3D runs, which are not force-free. However, for the high ${\mathcal M}_{\rm A}$ runs, the velocities generated in the pre-shock medium are two orders of magnitude smaller than turbulent velocities in the post-shock medium, and consequently are negligible. In the small ${\mathcal M}_{\rm A}$ runs, pre-existing velocities are not negligible, but in any case the turbulent dynamo is sub-dominant to compression amplification.

\begin{figure*}
  \begin{subfigure}[b]{0.48\textwidth}
    \centering
    \includegraphics[width=\textwidth]{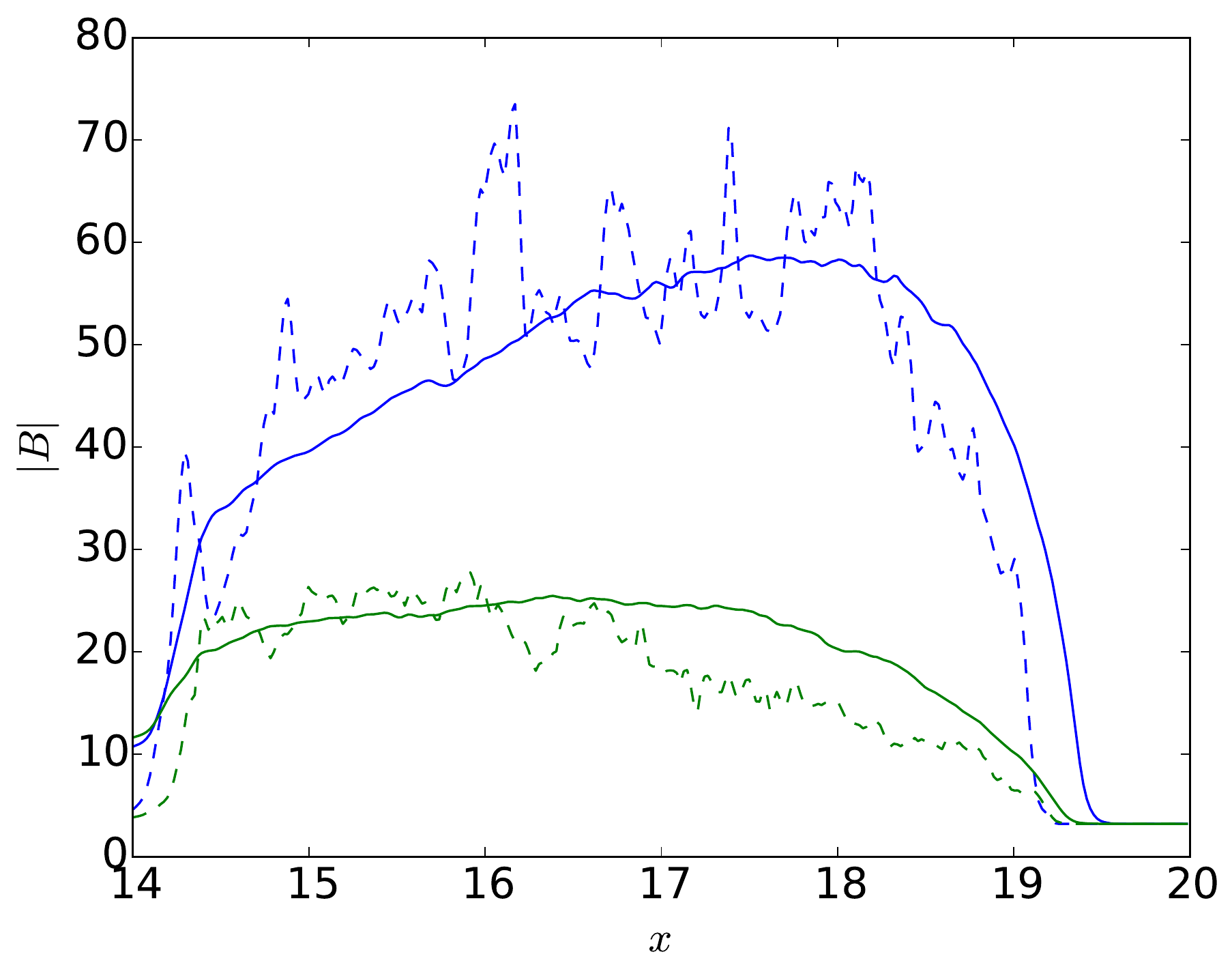}
    \caption{Magnetic field strength, $\mathcal{M}=133, \beta=1$}
    \label{fig:bmag_mach100_sn_2d3d}
  \end{subfigure}
  \begin{subfigure}[b]{0.48\textwidth}
    \centering
    \includegraphics[width=\textwidth]{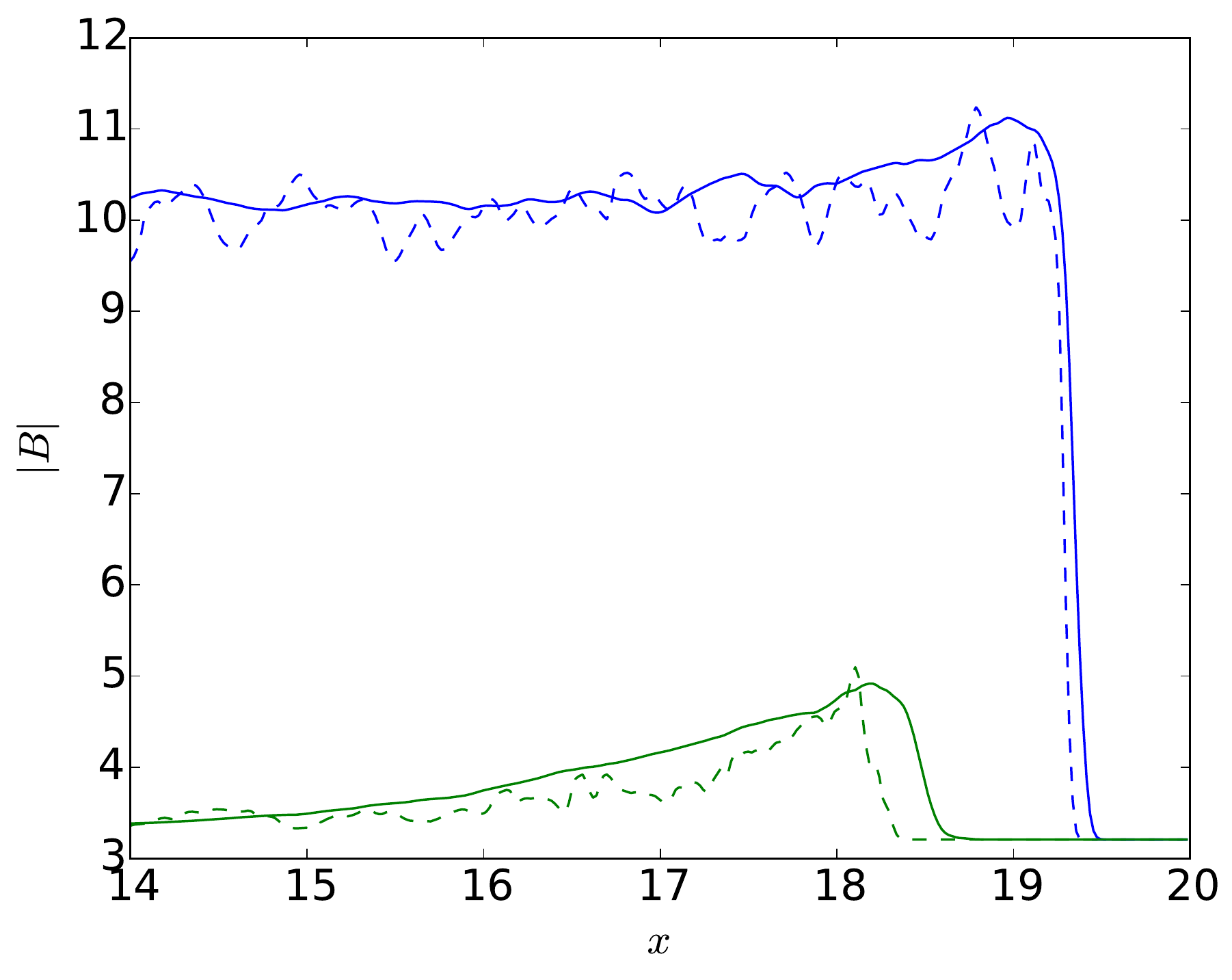}
    \caption{Magnetic field strength, $\mathcal{M}=4, \beta=1$}
    \label{fig:bmag_mach3_cluster_2d3d}
  \end{subfigure}
  \caption{2D vs. 3D $B$-field profiles with the resolution of $1024\times256$ ($1024\times256\times256$): dashed line -- 2D, solid line -- 3D; blue -- with initially perpendicular field, green -- with initially parallel field.}
  \label{fig:2d3d}
\end{figure*}

\subsection{Upstream field geometry; 2D versus 3D simulations}

We now examine the impact of upstream field geometry, and how simulation results differ in 2D and 3D. Fig. \ref{fig:2d3d} compares 2D and 3D results for initially perpendicular and parallel fields, for simulations matched to the same resolution. We see that there is good agreement between 2D and 3D simulations for perpendicular fields, while for parallel fields, differ slightly but remain broadly consistent.

In the plots the peak $B$-field amplification in the perpendicular and parallel cases differ by a factor of $\sim 2-3$. For an isotropic field, we have also found that amplification is a factor of $\sim \sqrt{2}/3 \sim 0.5$ of the Fig. \ref{fig:mach_macha_var} values. Note that no compressional growth is expected for a strictly parallel field in an undeformed shock; for the $\mathcal{M}=4$ compression dominated case, where only a weak turbulent dynamo operates, the difference in amplification can be understood in this light. What is more interesting is the much slower initial growth for the parallel field in the ${\mathcal{M} \sim 133}$ case, which is likely due to the fact that a parallel field is initially harder to deform and twist: the dominant velocity dispersion, in the $x$-direction, does not deform the field. The asymptotic maximum of $B$-field strength depends only on the shock Mach number and is for the stretching dominated high Mach number case, is eventually the same for perpendicular and parallel fields. However, for the parallel case, it reaches this maximum much further downstream (the plot ends when the field is still in the linear growth phase). These results suggest that both in the compression and stretching dominated cases, rapid amplification to strong $B$-fields indicated by strong radio emission immediately downstream of the shock require perpendicular fields. Along with the increased efficiency of cosmic ray acceleration at perpendicular shocks (e.g., \citet{giacalone05}), this potentially explains why bright supernova thin rims are only seen in certain regions (and in some cases, such as SN1006 and G1.9+3, have a bipolar symmetry).

\begin{figure*}
  \begin{subfigure}[b]{0.48\textwidth}
    \centering
    \includegraphics[width=\textwidth]{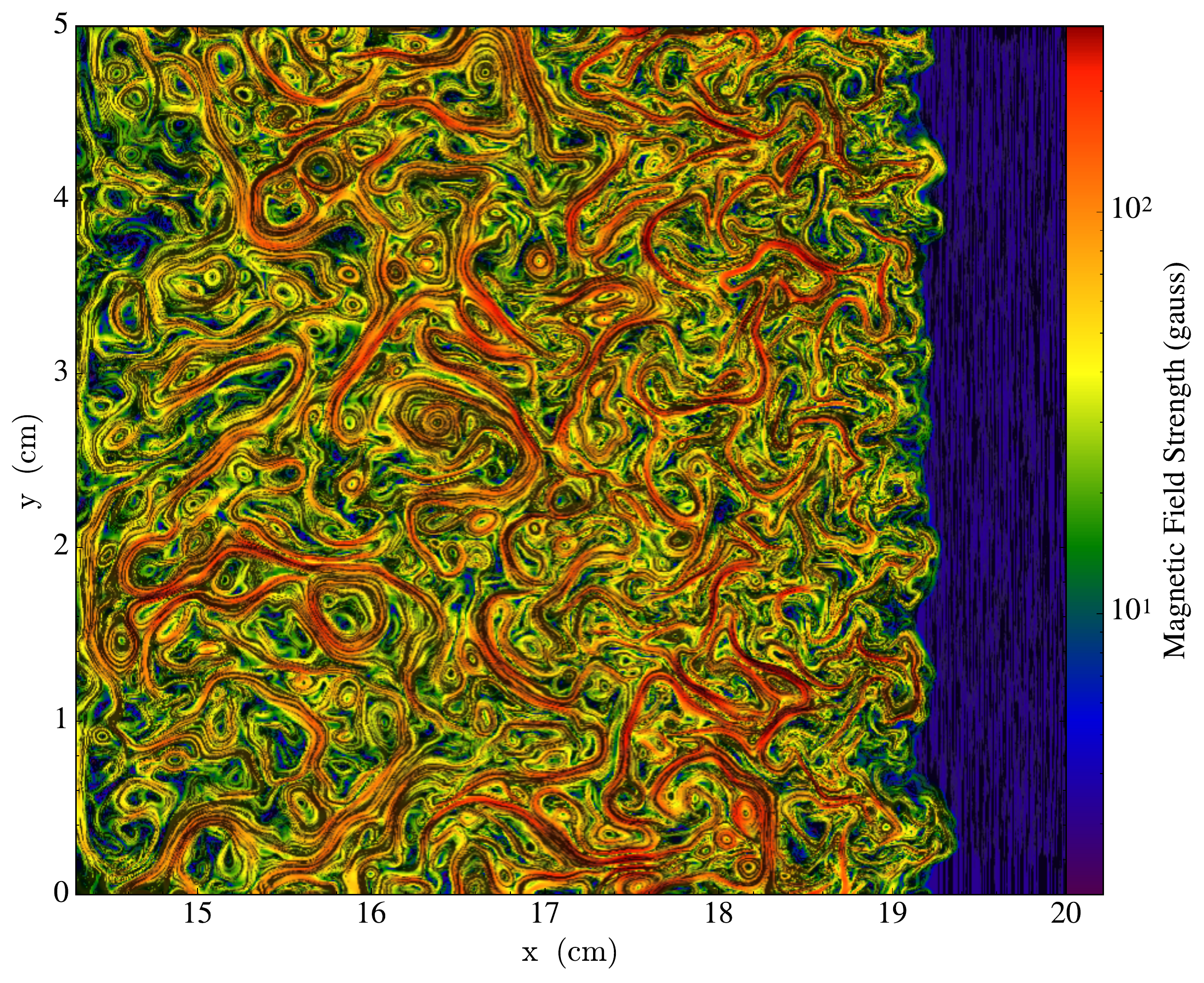}
    \caption{$\mathcal{M}=133, \beta=1$}
    \label{fig:2d_slice_sn}
  \end{subfigure}
  \begin{subfigure}[b]{0.48\textwidth}
    \centering
    \includegraphics[width=\textwidth]{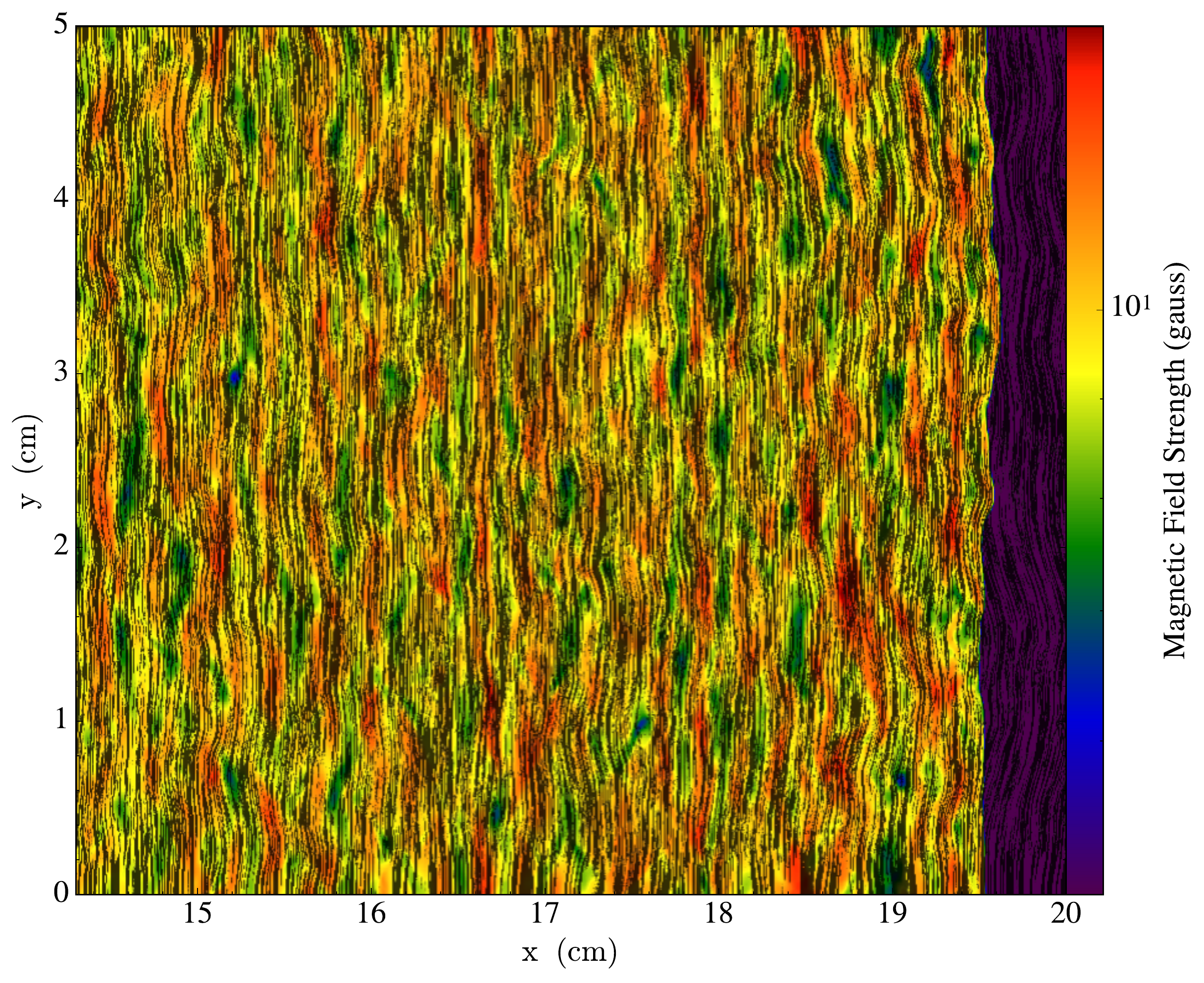}
    \caption{$\mathcal{M}=4, \beta=1$}
    \label{fig:2d_slice_cluster}
  \end{subfigure}
  \caption{Slice plots of magnetic field strength from 2D simulations with an initially perpendicular $B$-field, superposed with line integral convolution of magnetic vector field.}
  \label{fig:2d_slice}
\end{figure*}

\begin{figure*}
  \begin{subfigure}[b]{0.48\textwidth}
    \centering
    \includegraphics[width=\textwidth]{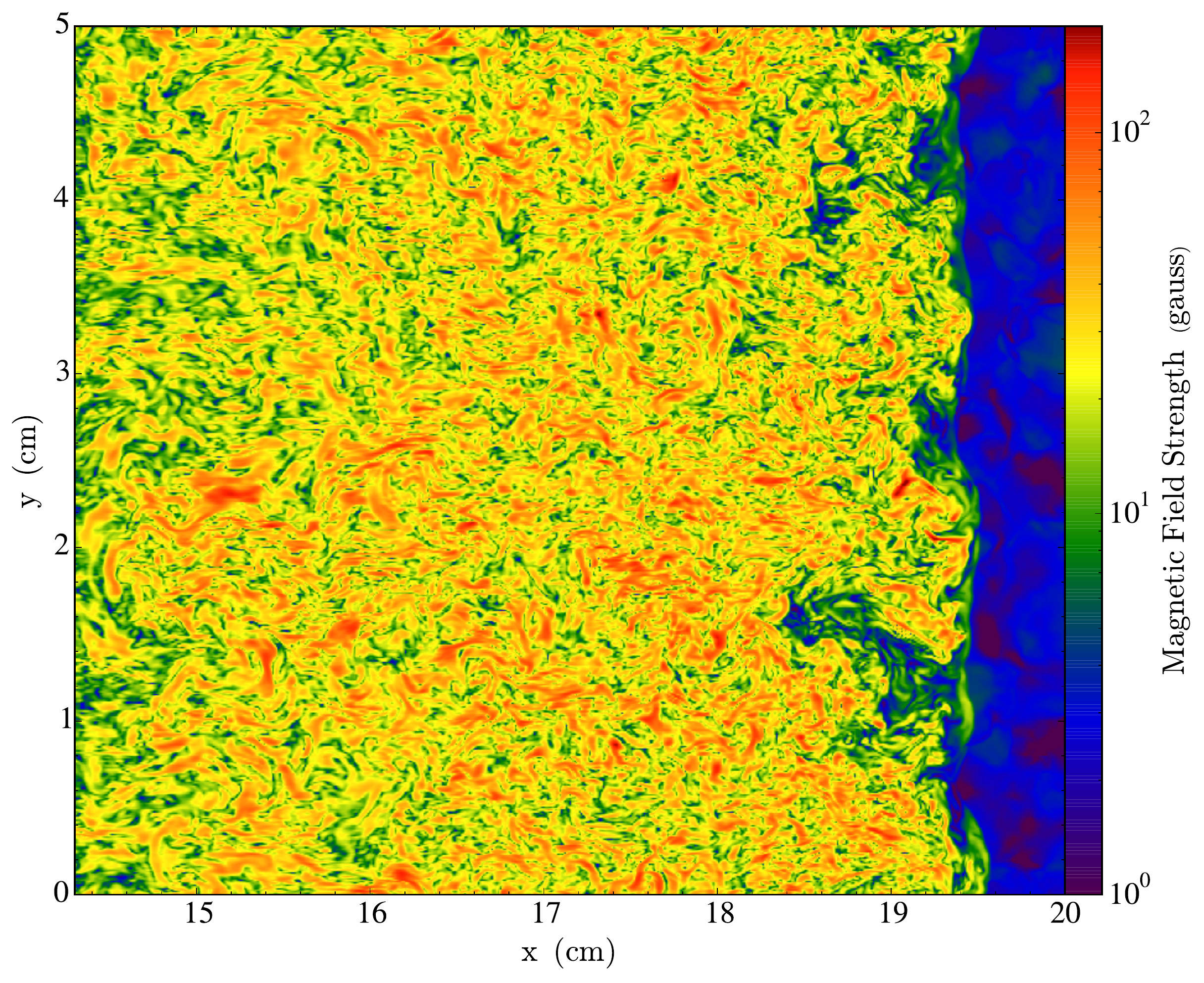}
    \caption{$\mathcal{M}=133, \beta=1$}
    \label{fig:slice_bmag_mach100_sn_3d_iso}
  \end{subfigure}
  \begin{subfigure}[b]{0.48\textwidth}
    \centering
    \includegraphics[width=\textwidth]{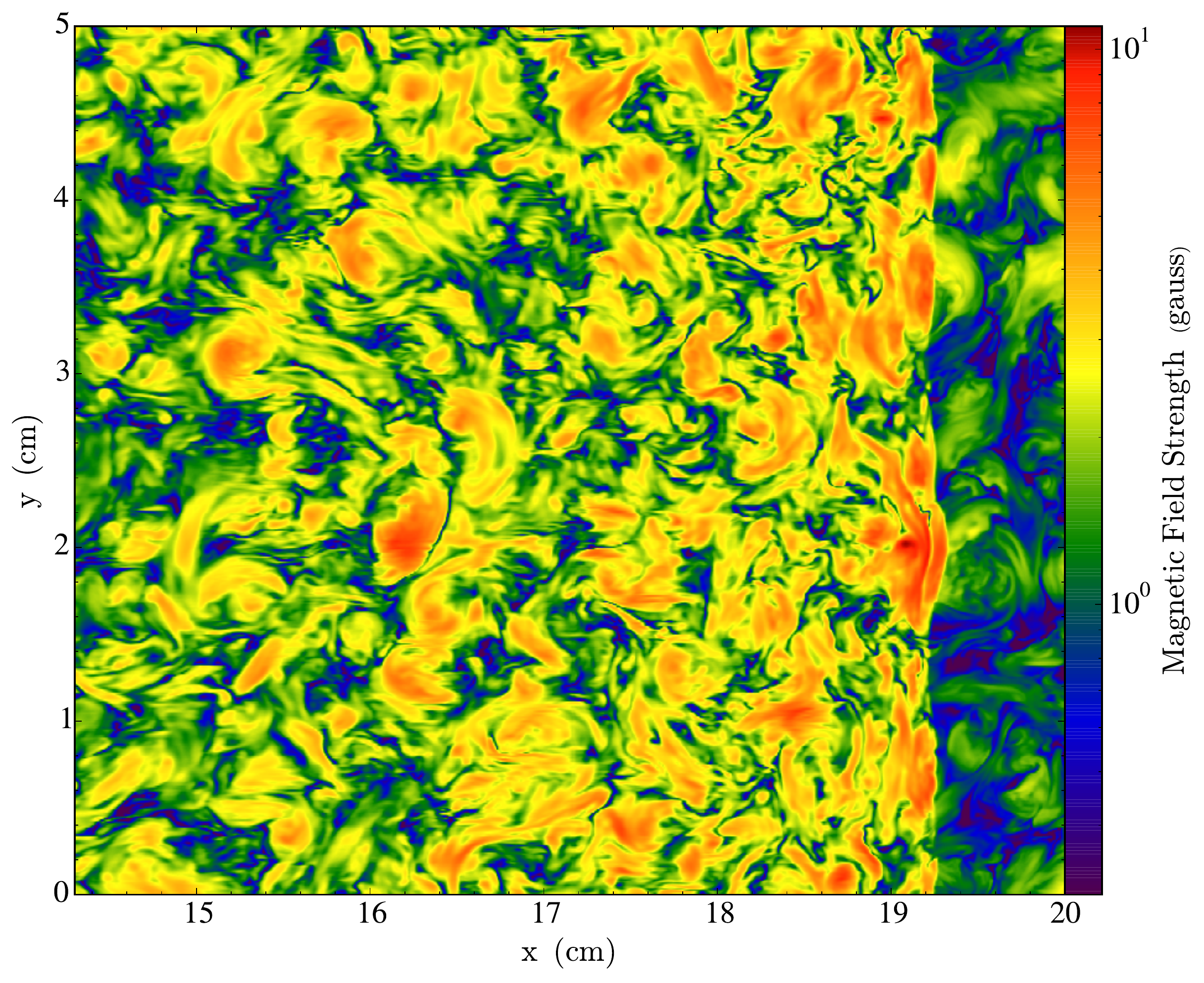}
    \caption{$\mathcal{M}=4, \beta=1$}
    \label{fig:slice_bmag_mach3_cluster_3d_iso}
  \end{subfigure}
  \caption{Slice plots of magnetic field strength from 3D simulations with initially isotropic magnetic field and resolution of $2048\times512\times512$.}
  \label{fig:slice_3d_iso}
\end{figure*}

Fig. \ref{fig:2d_slice} shows the slice plots of magnetic field from 2D simulations with an initially perpendicular field, superposed with line integral convolution of magnetic vector field. For strong shock case in Fig. \ref{fig:2d_slice_sn}, field lines are significantly stretched and the sizes of magnetic eddies are large. They increase downstream due to coalescence of eddies; the inverse cascade in 2D MHD means that eddies grow in size over time. In Fig. \ref{fig:slice_3d_iso}, we show the same quantities for a 3D simulation with an initially isotropic field. There is considerably more small scale field structure, in part due to the initially tangled field (a similar 3D calculation for a perpendicular field shows more large scale coherence), and in part due to the cascade to small scales in a 3D calculation. By contrast, in the low Mach number, compression dominated case, the field retains large scale coherence in both the 2D (perpendicular) and 3D (isotropic) cases.

\begin{figure*}
  \begin{subfigure}[b]{0.45\textwidth}
    \centering
    \includegraphics[width=\textwidth]{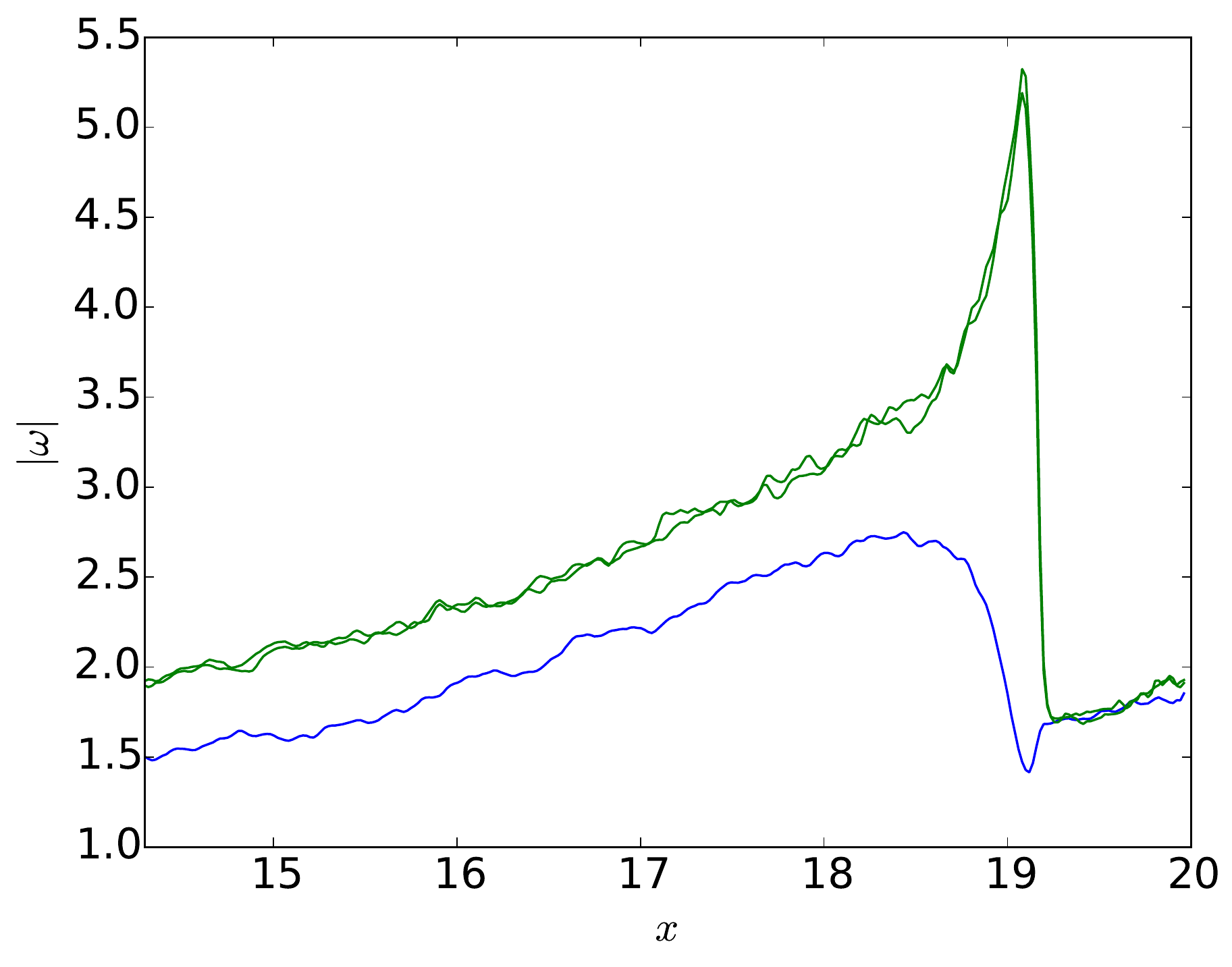}
    \caption{Radial profile of $|\omega|$, $\mathcal{M}=4, \beta=1$}
    \label{fig:3D_cluster_vorticity_anisotropy_lowbeta}
  \end{subfigure}
  \begin{subfigure}[b]{0.45\textwidth}
    \centering
    \includegraphics[width=\textwidth]{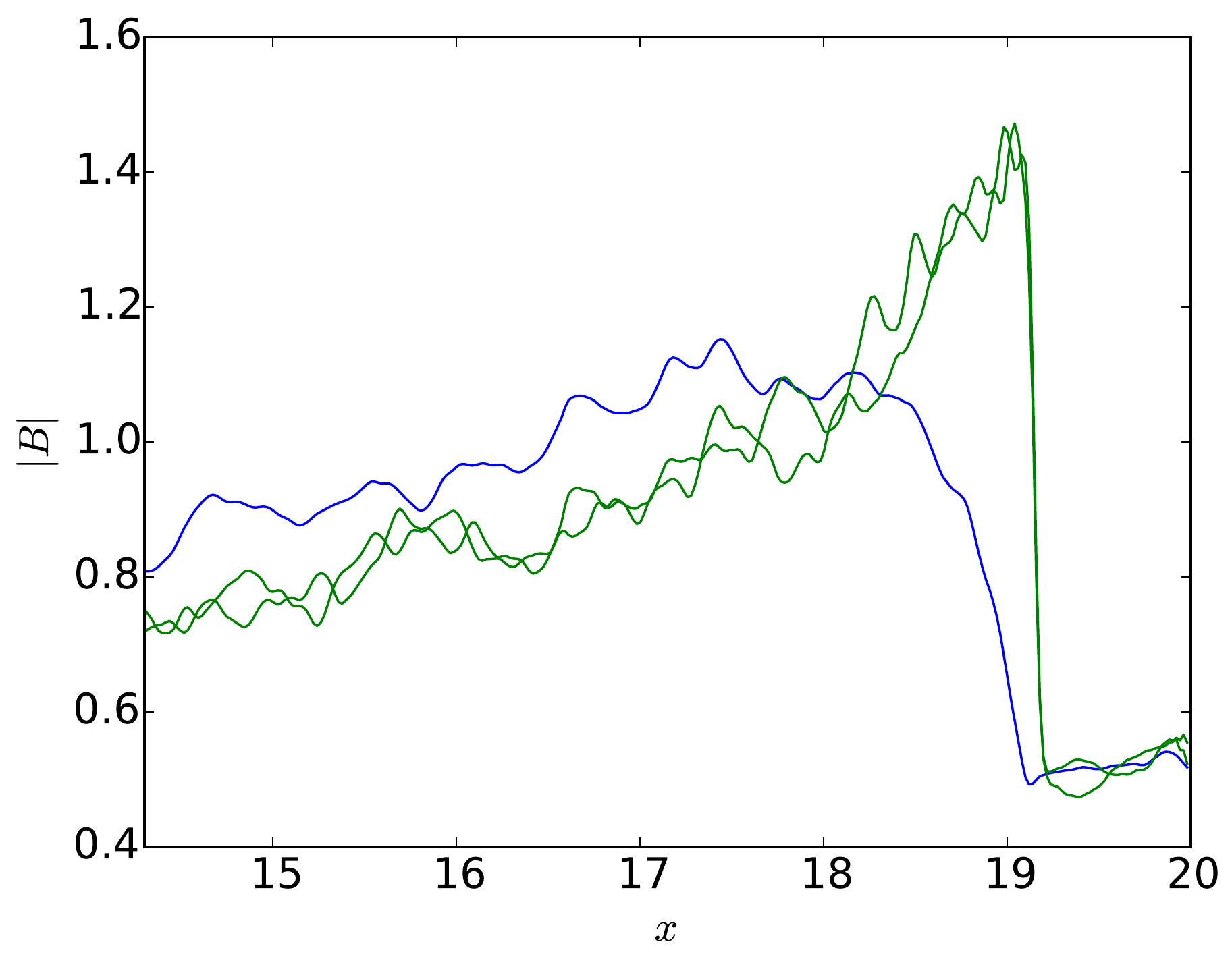}
    \caption{Radial profile of $|B|$, $\mathcal{M}=4, \beta=1$}
    \label{fig:3D_cluster_magnetic_anisotropy_lowbeta}
  \end{subfigure}
  \begin{subfigure}[b]{0.45\textwidth}
    \centering
    \includegraphics[width=\textwidth]{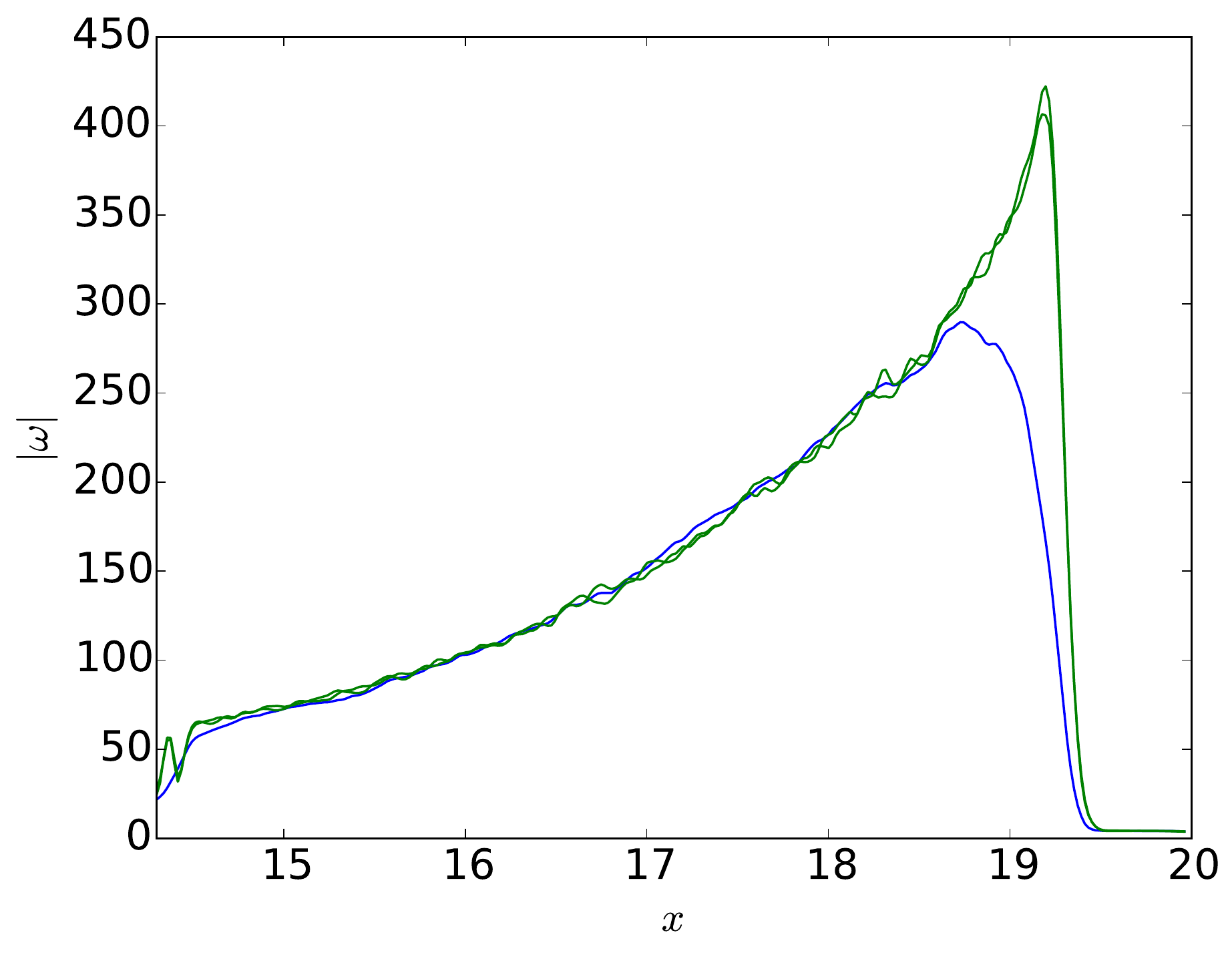}
    \caption{Radial profile of $|\omega|$, $\mathcal{M}=133, \beta=1$}
    \label{fig:3D_sn_vorticity_anisotropy}
  \end{subfigure}
  \begin{subfigure}[b]{0.45\textwidth}
    \centering
    \includegraphics[width=\textwidth]{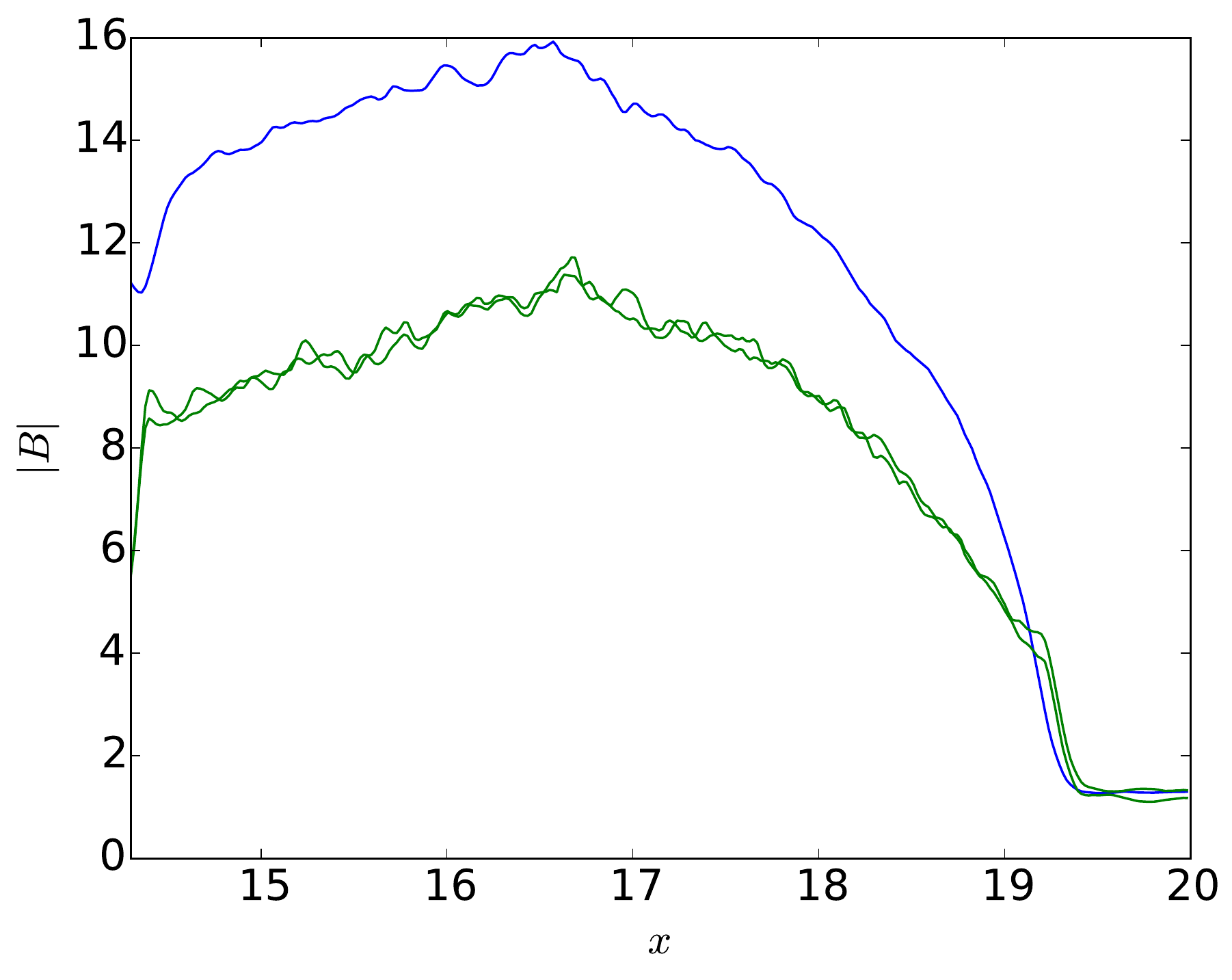}
    \caption{Radial profile of $|B|$, $\mathcal{M}=133, \beta=1$}
    \label{fig:3D_sn_magnetic_anisotropy}
  \end{subfigure}
  \caption{Radial profile of the absolute value of magnetic field and vorticity in three components from 3D simulations, with initially isotropic magnetic field. Blue line represents $x$ component, and green lines represent $y$ and $z$ components.}
  \label{fig:isotropy_3d_iso}
\end{figure*}

In the 2D strong shock case, the magnetic field in post-shock region quickly becomes isotropic regardless of the initial field configuration. However, in 3D, for a strong shock the downstream field clearly exhibits anisotropy, even with an initially isotropic field (see the far downstream region in Fig. \ref{fig:slice_3d_iso}, and see Fig. \ref{fig:3D_sn_magnetic_anisotropy}). This can be understood by considering the stretching term in the induction equation for the $B$-field: 
\begin{equation}
\left( \frac{\partial B_{i}}{\partial t} \right)_{\rm stretch} = \left( {\bm B} \cdot \nabla \right) {\bm v} = \left( B_{j} \frac{\partial}{\partial x_{j}} \right) v_{i}  
\label{eqn:stretch_vort} 
\end{equation}
where repeated indices imply summation, and comparing with vorticity: $\omega_{x}= \partial_{y} v_{z} - \partial_{z} v_{y}, \ \omega_{y} = \partial_{z} v_{x} - \partial_{x} v_{z}, \ \omega_{z} = \partial_{y} v_{x} - \partial_{x} v_{y}$. In 2D, there is only one non-zero component of vorticity, $\omega_{z}$, which contributes equally to $B_{x}, B_{y}$. Thus, field growth via the turbulent dynamo is isotropic in 2D. However, in 3D, all three components of vorticity are non-zero. Moreover, vorticity is anisotropic: $\omega_{x} < \omega_{y}, \omega_{z}$, with $\omega_{y} \approx \omega_{z}$. We can see this in Fig. \ref{fig:3D_sn_vorticity_anisotropy}. This makes sense: $\omega_{x}$ involves velocity gradients only in a plane parallel to the shock, and does not involve the dominant shear in the direction of shock propagation. Only after sufficient mixing, when velocity isotropy is achieved, do all 3 components of the vorticity equalize\footnote{There is an additional, subdominant effect: pre-existing vorticity (due to the non force-free $B$-fields) $\omega_{y,i} \omega_{z,i}$ are compressionally amplified, while $\omega_{x,i}$ is not}. Thus, field growth which involves $\omega_{y}, \omega_{z}$ instead of $\omega_{x}$ is going to be faster. Examining equation \ref{eqn:stretch_vort}, $\partial_{t} B_{x}$ has two such components, both involving the dominant motions in the x-direction: $B_{y} \partial_{y} v_{x}, B_{z} \partial_{z} v_{x}$, while $\partial_{t} B_{y}, \partial_{t} B_{z}$ each only have one component ($B_{x} \partial_{x} v_{y}$ and $B_{x} \partial_{x} v_{z}$, respectively). Thus, $B_{x}$ will be preferentially amplified when the turbulent dynamo operates. This radial bias has also been seen in previous 3D calculations \citep{inoue13,yang13}. This anisotropy in RMI generated $B$-fields can explain the radial $B$-field bias seen far downstream of the shock seen in polarization observations of supernova remnants. By contrast, the Rayleigh-Taylor instability \citep{jun96} only operates in the vicinity of the contact surface.

In short: if the field amplification is compression dominated, then the perpendicular field is preferentially amplified, in both the 2D and 3D calculations. For strong shocks, where turbulent amplification dominates, an anisotropic field with a parallel bias develops in the far-down stream, and the field quickly cascades to smaller scales due to the rapid development of turbulence.

\section{Conclusions}
\label{sect:conclusions} 

Our basic conclusion is that turbulent field amplification can be an important mechanism for generating the strong magnetic fields seen in the high Mach number shocks of supernovae, but its role is less clear in the low Mach number shocks of ICM radio relics. In supernovae, $\mathcal{M} \sim \mathcal{M}_\mathrm{A}\sim10^2$. From Fig. \ref{fig:mach_macha_var}, simulations in this regime suggest the amplification factor of over $100$ in maximum field strength, which is consistent with the fact that field strength in SNR can be magnified up a to $\sim100\ \mathrm{\mu G}$ from several $\mathrm{\mu G}$. Note also the field growth rates depend on magnetic geometry, and are strongest for perpendicular fields. Magnetic coherence lengths in the ISM are comparable to the size of supernova remnants and in some cases potentially larger. Thus, varying field geometry affects both cosmic-ray acceleration and field amplification and is plausibly responsible for the variation of synchrotron brightness around the remnant's shell. The polarization seen in the supernova remnant is also consistent with the radial $B$-field anisotropy we see far downstream of the shock when $B$-fields are amplified by RMI-generated turbulence. 

By contrast, in the ${\mathcal M} \sim 4$ shocks associated with radio relics in the ICM, the amplified magnetic field strength is $\sim  \mathrm{\mu G}$, while the pre-existing field strength is unknown. For $\beta \sim 1- 50$, $\mathcal{M}_{\mathrm A} \sim 3- 20$ (for $\mathcal{M} \sim 4$). From Fig. \ref{fig:stretch_compress_iso_vs_macha}, the effects of compression are comparable or larger than that of turbulence, with a maximum field amplification by a factor of $\sim 5$. Of course, it is important to note that since $B/B_{0} \propto {\mathcal M}_{\rm A}$, the post-shock $B$-field strength is approximately independent of its upstream value. Thus, the turbulence dynamo {\it{can}} explain observed field strengths. For very weak initial fields, the peak $B$-field can be significantly downstream of the shock, which is potentially relevant to the offset seen between X-ray and radio in some systems. However, compressional amplification is also consistent with the strong perpendicular anisotropy in the $B$-field behind the shock, as revealed by polarization measurements. This occurs even if the pre-shock $B$-field is isotropic. If instead the field were turbulently amplified, the resultant random field topology would be inconsistent with observations. This thus requires that the ambient ICM in the pre-shock field have field strength of order $\sim 0.2-1 \mu{\rm G}$, which is not implausible if the $B$-field field is in equipartition with turbulence (and the turbulent energy density is in turn a significant fraction of the thermal energy density). Such an explanation has been canonical for the fields seen in radio relics (e.g., \citet{iapichino12}).

Our other conclusions are as follows:
\begin{enumerate}
\item Shock amplification by RMI induced turbulence is proportional to the Alfv\'en Mach number: $B/B_{0} \propto {\mathcal M}_{\rm A}$, independent of the plasma $\beta$. Our main takeaway plot is Fig. \ref{fig:mach_macha_var}. As expected from dynamo theory, the $B$-field undergoes exponential growth followed by linear growth with time.

\item Turbulence (compression) dominates amplification at high (low) ${\mathcal M}_{\rm A}$, with a crossover at ${\mathcal M}_{\rm A} \sim 10$. Compression produces a post-shock field with a perpendicular bias, while turbulent amplification produces a downstream parallel bias. The parallel bias can be understood from the dominant velocity shear in the shock propagation direction creating anisotropic vorticity, and is only apparent in 3D simulations (fields are isotropic in 2D).

\item In a strong shock, turbulent and magnetic energy densities reach equipartition a distance of $\sim$ few $L_{\rm max}$ (where $L_{\rm max}$ is the outer scale of density fluctuations and induced turbulence) downstream of the shock. Turbulence decays due to transfer of energy to magnetic fields, and stabilization by magnetic tension.

\item Vorticity (which rises sharply at the shock and rapidly decays) is unconverged in our simulations, and increases with higher resolution. Fortunately, convergence requirements are much less stringent for $B$-fields, which are numerically converged even if vorticity is not. The asymptotic magnetic energy density is dominated by gas motion at large scales, while vorticity is dominated by small scale dynamics. However, the exponential phase of $B$-field amplification in the immediate postshock region is sensitive to resolution effects.

\item Above a critical threshold of $C \sim 1.5$, $B$-field amplification quickly asymptotes and depends only weakly on the gas clumping factor $C \equiv \langle \rho^{2} \rangle/\langle \rho \rangle^{2}$. This is fortunately, since the latter is quite uncertain.

\item For conditions in the ISM and ICM, we expect turbulence generated by the RMI to overwhelm compressional amplification of pre-existing turbulence (which should be unimportant in dynamo operation).

\item Field amplification is strongest for a perpendicular field; it is lower by a factor of $\sim 2.5$ for a parallel field, and $\sim 2$ for an isotropic field. Moreover, field growth is slower for a parallel field; we expect strong immediate postshock fields to be associated with perpendicular pre-shock fields.

\end{enumerate} 

\section{Acknowledgements} 
We thank Mike McCourt for helpful conversations and insightful comments on a draft. SJ and SPO are supported by NASA grant NNX12AG73G and NNX15AK81G. MR acknowledges NSF grant NSF 1008454 and NASA ATP12-0017. This research has used the Extreme Science and Engineering Discovery Environment (XSEDE allocations TG-AST140058 and TG-AST140086). We have made use of NASA's Astrophysics Data System and the yt astrophysics analysis software suite \citep{turk2010yt}.

\bibliographystyle{mnras}
\bibliography{master_references}

\begin{thebibliography}{}
\makeatletter
\relax
\def\mn@urlcharsother{\let\do\@makeother \do\$\do\&\do\#\do\^\do\_\do\%\do\~}
\def\mn@doi{\begingroup\mn@urlcharsother \@ifnextchar [ {\mn@doi@}
  {\mn@doi@[]}}
\def\mn@doi@[#1]#2{\def\@tempa{#1}\ifx\@tempa\@empty \href
  {http://dx.doi.org/#2} {doi:#2}\else \href {http://dx.doi.org/#2} {#1}\fi
  \endgroup}
\def\mn@eprint#1#2{\mn@eprint@#1:#2::\@nil}
\def\mn@eprint@arXiv#1{\href {http://arxiv.org/abs/#1} {{\tt arXiv:#1}}}
\def\mn@eprint@dblp#1{\href {http://dblp.uni-trier.de/rec/bibtex/#1.xml}
  {dblp:#1}}
\def\mn@eprint@#1:#2:#3:#4\@nil{\def\@tempa {#1}\def\@tempb {#2}\def\@tempc
  {#3}\ifx \@tempc \@empty \let \@tempc \@tempb \let \@tempb \@tempa \fi \ifx
  \@tempb \@empty \def\@tempb {arXiv}\fi \@ifundefined
  {mn@eprint@\@tempb}{\@tempb:\@tempc}{\expandafter \expandafter \csname
  mn@eprint@\@tempb\endcsname \expandafter{\@tempc}}}

\bibitem[\protect\citeauthoryear{{Armstrong}, {Rickett}  \&
  {Spangler}}{{Armstrong} et~al.}{1995}]{armstrong95}
{Armstrong} J.~W.,  {Rickett} B.~J.,   {Spangler} S.~R.,  1995, \mn@doi [\apj]
  {10.1086/175515}, \href {http://adsabs.harvard.edu/abs/1995ApJ...443..209A}
  {443, 209}

\bibitem[\protect\citeauthoryear{Aspden, Nikiforakis, Dalziel  \& Bell}{Aspden
  et~al.}{2009}]{aspden09}
Aspden A.,  Nikiforakis N.,  Dalziel S.,   Bell J.,  2009, Communications in
  Applied Mathematics and Computational Science, 3, 103

\bibitem[\protect\citeauthoryear{{Balsara} \& {Kim}}{{Balsara} \&
  {Kim}}{2005}]{balsara05}
{Balsara} D.~S.,  {Kim} J.,  2005, \mn@doi [\apj] {10.1086/452626}, \href
  {http://adsabs.harvard.edu/abs/2005ApJ...634..390B} {634, 390}

\bibitem[\protect\citeauthoryear{{Bamba}, {Yamazaki}, {Yoshida}, {Terasawa}  \&
  {Koyama}}{{Bamba} et~al.}{2005}]{bamba05}
{Bamba} A.,  {Yamazaki} R.,  {Yoshida} T.,  {Terasawa} T.,   {Koyama} K.,
  2005, \mn@doi [\apj] {10.1086/427620}, \href
  {http://adsabs.harvard.edu/abs/2005ApJ...621..793B} {621, 793}

\bibitem[\protect\citeauthoryear{{Battaglia}, {Pfrommer}, {Sievers}, {Bond}  \&
  {En{\ss}lin}}{{Battaglia} et~al.}{2009}]{battaglia09}
{Battaglia} N.,  {Pfrommer} C.,  {Sievers} J.~L.,  {Bond} J.~R.,   {En{\ss}lin}
  T.~A.,  2009, \mn@doi [\mnras] {10.1111/j.1365-2966.2008.14136.x}, \href
  {http://adsabs.harvard.edu/abs/2009MNRAS.393.1073B} {393, 1073}

\bibitem[\protect\citeauthoryear{{Battaglia}, {Bond}, {Pfrommer}  \&
  {Sievers}}{{Battaglia} et~al.}{2012}]{battaglia12}
{Battaglia} N.,  {Bond} J.~R.,  {Pfrommer} C.,   {Sievers} J.~L.,  2012,
  \mn@doi [\apj] {10.1088/0004-637X/758/2/74}, \href
  {http://ads.nao.ac.jp/abs/2012ApJ...758...74B} {758, 74}

\bibitem[\protect\citeauthoryear{{Battaglia}, {Bond}, {Pfrommer}  \&
  {Sievers}}{{Battaglia} et~al.}{2013}]{battaglia13}
{Battaglia} N.,  {Bond} J.~R.,  {Pfrommer} C.,   {Sievers} J.~L.,  2013,
  \mn@doi [\apj] {10.1088/0004-637X/777/2/123}, \href
  {http://adsabs.harvard.edu/abs/2013ApJ...777..123B} {777, 123}

\bibitem[\protect\citeauthoryear{{Bell}}{{Bell}}{2004}]{bell04}
{Bell} A.~R.,  2004, \mn@doi [\mnras] {10.1111/j.1365-2966.2004.08097.x}, \href
  {http://adsabs.harvard.edu/abs/2004MNRAS.353..550B} {353, 550}

\bibitem[\protect\citeauthoryear{{Beresnyak}, {Jones}  \&
  {Lazarian}}{{Beresnyak} et~al.}{2009}]{beresnyak09}
{Beresnyak} A.,  {Jones} T.~W.,   {Lazarian} A.,  2009, \mn@doi [\apj]
  {10.1088/0004-637X/707/2/1541}, \href
  {http://adsabs.harvard.edu/abs/2009ApJ...707.1541B} {707, 1541}

\bibitem[\protect\citeauthoryear{{Bonafede}, {Feretti}, {Murgia}, {Govoni},
  {Giovannini}, {Dallacasa}, {Dolag}  \& {Taylor}}{{Bonafede}
  et~al.}{2010}]{bonafede10}
{Bonafede} A.,  {Feretti} L.,  {Murgia} M.,  {Govoni} F.,  {Giovannini} G.,
  {Dallacasa} D.,  {Dolag} K.,   {Taylor} G.~B.,  2010, \mn@doi [\aap]
  {10.1051/0004-6361/200913696}, \href
  {http://adsabs.harvard.edu/abs/2010A%26A...513A..30B} {513, A30}

\bibitem[\protect\citeauthoryear{{Br{\"u}ggen}}{{Br{\"u}ggen}}{2013}]{bruggen13}
{Br{\"u}ggen} M.,  2013, \mn@doi [\mnras] {10.1093/mnras/stt1566}, \href
  {http://adsabs.harvard.edu/abs/2013MNRAS.436..294B} {436, 294}

\bibitem[\protect\citeauthoryear{{Br{\"u}ggen}, {Bykov}, {Ryu}  \&
  {R{\"o}ttgering}}{{Br{\"u}ggen} et~al.}{2012}]{bruggen12}
{Br{\"u}ggen} M.,  {Bykov} A.,  {Ryu} D.,   {R{\"o}ttgering} H.,  2012, \mn@doi
  [\ssr] {10.1007/s11214-011-9785-9}, \href
  {http://adsabs.harvard.edu/abs/2012SSRv..166..187B} {166, 187}

\bibitem[\protect\citeauthoryear{{Brunetti} \& {Jones}}{{Brunetti} \&
  {Jones}}{2014}]{brunetti14}
{Brunetti} G.,  {Jones} T.~W.,  2014, \mn@doi [International Journal of Modern
  Physics D] {10.1142/S0218271814300079}, \href
  {http://adsabs.harvard.edu/abs/2014IJMPD..2330007B} {23, 30007}

\bibitem[\protect\citeauthoryear{{Cho}, {Vishniac}, {Beresnyak}, {Lazarian}  \&
  {Ryu}}{{Cho} et~al.}{2009}]{cho09}
{Cho} J.,  {Vishniac} E.~T.,  {Beresnyak} A.,  {Lazarian} A.,   {Ryu} D.,
  2009, \mn@doi [\apj] {10.1088/0004-637X/693/2/1449}, \href
  {http://adsabs.harvard.edu/abs/2009ApJ...693.1449C} {693, 1449}

\bibitem[\protect\citeauthoryear{{DeLaney}, {Koralesky}, {Rudnick}  \&
  {Dickel}}{{DeLaney} et~al.}{2002}]{delaney02}
{DeLaney} T.,  {Koralesky} B.,  {Rudnick} L.,   {Dickel} J.~R.,  2002, \mn@doi
  [\apj] {10.1086/343787}, \href
  {http://adsabs.harvard.edu/abs/2002ApJ...580..914D} {580, 914}

\bibitem[\protect\citeauthoryear{{Dickel}, {van Breugel}  \& {Strom}}{{Dickel}
  et~al.}{1991}]{dickel91}
{Dickel} J.~R.,  {van Breugel} W.~J.~M.,   {Strom} R.~G.,  1991, \mn@doi [\aj]
  {10.1086/115837}, \href {http://adsabs.harvard.edu/abs/1991AJ....101.2151D}
  {101, 2151}

\bibitem[\protect\citeauthoryear{{Dolag}, {Borgani}, {Murante}  \&
  {Springel}}{{Dolag} et~al.}{2009}]{dolag09}
{Dolag} K.,  {Borgani} S.,  {Murante} G.,   {Springel} V.,  2009, \mn@doi
  [\mnras] {10.1111/j.1365-2966.2009.15034.x}, \href
  {http://adsabs.harvard.edu/abs/2009MNRAS.399..497D} {399, 497}

\bibitem[\protect\citeauthoryear{{Drury} \& {Downes}}{{Drury} \&
  {Downes}}{2012}]{drury12}
{Drury} L.~O.,  {Downes} T.~P.,  2012, \mn@doi [\mnras]
  {10.1111/j.1365-2966.2012.22106.x}, \href
  {http://adsabs.harvard.edu/abs/2012MNRAS.427.2308D} {427, 2308}

\bibitem[\protect\citeauthoryear{{Dubois} \& {Teyssier}}{{Dubois} \&
  {Teyssier}}{2008}]{dubois08}
{Dubois} Y.,  {Teyssier} R.,  2008, \mn@doi [\aap]
  {10.1051/0004-6361:200809513}, \href
  {http://adsabs.harvard.edu/abs/2008A%26A...482L..13D} {482, L13}

\bibitem[\protect\citeauthoryear{{Elmegreen} \& {Scalo}}{{Elmegreen} \&
  {Scalo}}{2004}]{elmegreen04}
{Elmegreen} B.~G.,  {Scalo} J.,  2004, \mn@doi [\araa]
  {10.1146/annurev.astro.41.011802.094859}, \href
  {http://adsabs.harvard.edu/abs/2004ARA%26A..42..211E} {42, 211}

\bibitem[\protect\citeauthoryear{{Federrath}, {Roman-Duval}, {Klessen},
  {Schmidt}  \& {Mac Low}}{{Federrath} et~al.}{2010}]{federrath10}
{Federrath} C.,  {Roman-Duval} J.,  {Klessen} R.~S.,  {Schmidt} W.,   {Mac Low}
  M.-M.,  2010, \mn@doi [\aap] {10.1051/0004-6361/200912437}, \href
  {http://adsabs.harvard.edu/abs/2010A%26A...512A..81F} {512, A81}

\bibitem[\protect\citeauthoryear{{Ferrari}, {Govoni}, {Schindler}, {Bykov}  \&
  {Rephaeli}}{{Ferrari} et~al.}{2008}]{ferrari08}
{Ferrari} C.,  {Govoni} F.,  {Schindler} S.,  {Bykov} A.~M.,   {Rephaeli} Y.,
  2008, \mn@doi [\ssr] {10.1007/s11214-008-9311-x}, \href
  {http://adsabs.harvard.edu/abs/2008SSRv..134...93F} {134, 93}

\bibitem[\protect\citeauthoryear{{Fraschetti}}{{Fraschetti}}{2013}]{fraschetti13}
{Fraschetti} F.,  2013, \mn@doi [\apj] {10.1088/0004-637X/770/2/84}, \href
  {http://adsabs.harvard.edu/abs/2013ApJ...770...84F} {770, 84}

\bibitem[\protect\citeauthoryear{{Fryxell} et~al.,}{{Fryxell}
  et~al.}{2000}]{fryxell00}
{Fryxell} B.,  et~al., 2000, \mn@doi [\apjs] {10.1086/317361}, \href
  {http://adsabs.harvard.edu/abs/2000ApJS..131..273F} {131, 273}

\bibitem[\protect\citeauthoryear{{Furlanetto} \& {Loeb}}{{Furlanetto} \&
  {Loeb}}{2001}]{furlanetto01}
{Furlanetto} S.~R.,  {Loeb} A.,  2001, \mn@doi [\apj] {10.1086/321630}, \href
  {http://adsabs.harvard.edu/abs/2001ApJ...556..619F} {556, 619}

\bibitem[\protect\citeauthoryear{{Gaspari} \& {Churazov}}{{Gaspari} \&
  {Churazov}}{2013}]{gaspari13a}
{Gaspari} M.,  {Churazov} E.,  2013, \mn@doi [\aap]
  {10.1051/0004-6361/201322295}, \href
  {http://adsabs.harvard.edu/abs/2013A%26A...559A..78G} {559, A78}

\bibitem[\protect\citeauthoryear{{Giacalone}}{{Giacalone}}{2005}]{giacalone05}
{Giacalone} J.,  2005, \mn@doi [\apjl] {10.1086/432510}, \href
  {http://adsabs.harvard.edu/abs/2005ApJ...628L..37G} {628, L37}

\bibitem[\protect\citeauthoryear{{Giacalone} \& {Jokipii}}{{Giacalone} \&
  {Jokipii}}{2007}]{giacalone07}
{Giacalone} J.,  {Jokipii} J.~R.,  2007, \mn@doi [\apjl] {10.1086/519994},
  \href {http://adsabs.harvard.edu/abs/2007ApJ...663L..41G} {663, L41}

\bibitem[\protect\citeauthoryear{{Goldreich} \& {Sridhar}}{{Goldreich} \&
  {Sridhar}}{1995}]{goldreich95}
{Goldreich} P.,  {Sridhar} S.,  1995, \mn@doi [\apj] {10.1086/175121}, \href
  {http://adsabs.harvard.edu/abs/1995ApJ...438..763G} {438, 763}

\bibitem[\protect\citeauthoryear{{Guo}, {Li}, {Li}, {Giacalone}, {Jokipii}  \&
  {Li}}{{Guo} et~al.}{2012}]{guo12}
{Guo} F.,  {Li} S.,  {Li} H.,  {Giacalone} J.,  {Jokipii} J.~R.,   {Li} D.,
  2012, \mn@doi [\apj] {10.1088/0004-637X/747/2/98}, \href
  {http://adsabs.harvard.edu/abs/2012ApJ...747...98G} {747, 98}

\bibitem[\protect\citeauthoryear{{Hoeft}, {Br{\"u}ggen}, {Yepes},
  {Gottl{\"o}ber}  \& {Schwope}}{{Hoeft} et~al.}{2008}]{hoeft08}
{Hoeft} M.,  {Br{\"u}ggen} M.,  {Yepes} G.,  {Gottl{\"o}ber} S.,   {Schwope}
  A.,  2008, \mn@doi [\mnras] {10.1111/j.1365-2966.2008.13955.x}, \href
  {http://adsabs.harvard.edu/abs/2008MNRAS.391.1511H} {391, 1511}

\bibitem[\protect\citeauthoryear{{Iapichino} \& {Br{\"u}ggen}}{{Iapichino} \&
  {Br{\"u}ggen}}{2012}]{iapichino12}
{Iapichino} L.,  {Br{\"u}ggen} M.,  2012, \mn@doi [\mnras]
  {10.1111/j.1365-2966.2012.21084.x}, \href
  {http://adsabs.harvard.edu/abs/2012MNRAS.423.2781I} {423, 2781}

\bibitem[\protect\citeauthoryear{{Inoue}, {Yamazaki}  \& {Inutsuka}}{{Inoue}
  et~al.}{2009}]{inoue09}
{Inoue} T.,  {Yamazaki} R.,   {Inutsuka} S.-i.,  2009, \mn@doi [\apj]
  {10.1088/0004-637X/695/2/825}, \href
  {http://adsabs.harvard.edu/abs/2009ApJ...695..825I} {695, 825}

\bibitem[\protect\citeauthoryear{{Inoue}, {Shimoda}, {Ohira}  \&
  {Yamazaki}}{{Inoue} et~al.}{2013}]{inoue13}
{Inoue} T.,  {Shimoda} J.,  {Ohira} Y.,   {Yamazaki} R.,  2013, \mn@doi [\apjl]
  {10.1088/2041-8205/772/2/L20}, \href
  {http://adsabs.harvard.edu/abs/2013ApJ...772L..20I} {772, L20}

\bibitem[\protect\citeauthoryear{{Jun} \& {Norman}}{{Jun} \&
  {Norman}}{1996}]{jun96}
{Jun} B.-I.,  {Norman} M.~L.,  1996, \mn@doi [\apj] {10.1086/178059}, \href
  {http://adsabs.harvard.edu/abs/1996ApJ...472..245J} {472, 245}

\bibitem[\protect\citeauthoryear{{Kevlahan}}{{Kevlahan}}{1997}]{kevlahan97}
{Kevlahan} N.~K.-R.,  1997, Journal of Fluid Mechanics, \href
  {http://adsabs.harvard.edu/abs/1997JFM...341..371K} {341, 371}

\bibitem[\protect\citeauthoryear{{Kevlahan} \& {Pudritz}}{{Kevlahan} \&
  {Pudritz}}{2009}]{kevlahan09}
{Kevlahan} N.,  {Pudritz} R.~E.,  2009, \mn@doi [\apj]
  {10.1088/0004-637X/702/1/39}, \href
  {http://adsabs.harvard.edu/abs/2009ApJ...702...39K} {702, 39}

\bibitem[\protect\citeauthoryear{{Kritsuk}, {Norman}, {Padoan}  \&
  {Wagner}}{{Kritsuk} et~al.}{2007}]{kritsuk07}
{Kritsuk} A.~G.,  {Norman} M.~L.,  {Padoan} P.,   {Wagner} R.,  2007, \mn@doi
  [\apj] {10.1086/519443}, \href
  {http://adsabs.harvard.edu/abs/2007ApJ...665..416K} {665, 416}

\bibitem[\protect\citeauthoryear{{Kulsrud}}{{Kulsrud}}{2005}]{kulsrud05}
{Kulsrud} R.~M.,  2005, {Plasma physics for astrophysics}.
Princeton University Press

\bibitem[\protect\citeauthoryear{{Lazarian} \& {Pogosyan}}{{Lazarian} \&
  {Pogosyan}}{2000}]{lazarian00}
{Lazarian} A.,  {Pogosyan} D.,  2000, \mn@doi [\apj] {10.1086/309040}, \href
  {http://adsabs.harvard.edu/abs/2000ApJ...537..720L} {537, 720}

\bibitem[\protect\citeauthoryear{{Lecoanet} et~al.,}{{Lecoanet}
  et~al.}{2016}]{lecoanet16}
{Lecoanet} D.,  et~al., 2016, \mn@doi [\mnras] {10.1093/mnras/stv2564}, \href
  {http://adsabs.harvard.edu/abs/2016MNRAS.455.4274L} {455, 4274}

\bibitem[\protect\citeauthoryear{Lee}{Lee}{2013}]{lee2013solution}
Lee D.,  2013, Journal of Computational Physics, 243, 269

\bibitem[\protect\citeauthoryear{{Leroy} et~al.,}{{Leroy}
  et~al.}{2013}]{leroy13}
{Leroy} A.~K.,  et~al., 2013, \mn@doi [\apjl] {10.1088/2041-8205/769/1/L12},
  \href {http://adsabs.harvard.edu/abs/2013ApJ...769L..12L} {769, L12}

\bibitem[\protect\citeauthoryear{Meinecke et~al.,}{Meinecke
  et~al.}{2014}]{meinecke14}
Meinecke J.,  et~al., 2014, Nature Physics

\bibitem[\protect\citeauthoryear{{Pohl}, {Yan}  \& {Lazarian}}{{Pohl}
  et~al.}{2005}]{pohl05}
{Pohl} M.,  {Yan} H.,   {Lazarian} A.,  2005, \mn@doi [\apjl] {10.1086/431902},
  \href {http://adsabs.harvard.edu/abs/2005ApJ...626L.101P} {626, L101}

\bibitem[\protect\citeauthoryear{{Porter}, {Jones}  \& {Ryu}}{{Porter}
  et~al.}{2015}]{porter15}
{Porter} D.~H.,  {Jones} T.~W.,   {Ryu} D.,  2015, \mn@doi [\apj]
  {10.1088/0004-637X/810/2/93}, \href
  {http://adsabs.harvard.edu/abs/2015ApJ...810...93P} {810, 93}

\bibitem[\protect\citeauthoryear{{Radice}, {Couch}  \& {Ott}}{{Radice}
  et~al.}{2015}]{radice15}
{Radice} D.,  {Couch} S.~M.,   {Ott} C.~D.,  2015, \mn@doi [Computational
  Astrophysics and Cosmology] {10.1186/s40668-015-0011-0}, \href
  {http://adsabs.harvard.edu/abs/2015ComAC...2....7R} {2, 7}

\bibitem[\protect\citeauthoryear{Ressler, Katsuda, Reynolds, Long, Petre,
  Williams  \& Winkler}{Ressler et~al.}{2014}]{ressler2014magnetic}
Ressler S.~M.,  Katsuda S.,  Reynolds S.~P.,  Long K.~S.,  Petre R.,  Williams
  B.~J.,   Winkler P.~F.,  2014, The Astrophysical Journal, 790, 85

\bibitem[\protect\citeauthoryear{{Reynoso}, {Hughes}  \& {Moffett}}{{Reynoso}
  et~al.}{2013}]{reynoso13}
{Reynoso} E.~M.,  {Hughes} J.~P.,   {Moffett} D.~A.,  2013, \mn@doi [\aj]
  {10.1088/0004-6256/145/4/104}, \href
  {http://adsabs.harvard.edu/abs/2013AJ....145..104R} {145, 104}

\bibitem[\protect\citeauthoryear{Richtmyer}{Richtmyer}{1960}]{richtmyer60}
Richtmyer R.~D.,  1960, \mn@doi [Communications on Pure and Applied
  Mathematics] {10.1002/cpa.3160130207}, 13, 297

\bibitem[\protect\citeauthoryear{{Ridge} et~al.,}{{Ridge}
  et~al.}{2006}]{ridge06}
{Ridge} N.~A.,  et~al., 2006, \mn@doi [\aj] {10.1086/503704}, \href
  {http://adsabs.harvard.edu/abs/2006AJ....131.2921R} {131, 2921}

\bibitem[\protect\citeauthoryear{{Riquelme} \& {Spitkovsky}}{{Riquelme} \&
  {Spitkovsky}}{2009}]{riquelme09}
{Riquelme} M.~A.,  {Spitkovsky} A.,  2009, \mn@doi [\apj]
  {10.1088/0004-637X/694/1/626}, \href
  {http://adsabs.harvard.edu/abs/2009ApJ...694..626R} {694, 626}

\bibitem[\protect\citeauthoryear{{Riquelme} \& {Spitkovsky}}{{Riquelme} \&
  {Spitkovsky}}{2010}]{riquelme10}
{Riquelme} M.~A.,  {Spitkovsky} A.,  2010, \mn@doi [\apj]
  {10.1088/0004-637X/717/2/1054}, \href
  {http://adsabs.harvard.edu/abs/2010ApJ...717.1054R} {717, 1054}

\bibitem[\protect\citeauthoryear{{Roberts}}{{Roberts}}{2010}]{Roberts10}
{Roberts} D.~A.,  2010, \mn@doi [Journal of Geophysical Research (Space
  Physics)] {10.1029/2009JA015120}, \href
  {http://adsabs.harvard.edu/abs/2010JGRA..11512101R} {115, A12101}

\bibitem[\protect\citeauthoryear{{Sano}, {Nishihara}, {Matsuoka}  \&
  {Inoue}}{{Sano} et~al.}{2012}]{sano12}
{Sano} T.,  {Nishihara} K.,  {Matsuoka} C.,   {Inoue} T.,  2012, \mn@doi [\apj]
  {10.1088/0004-637X/758/2/126}, \href
  {http://adsabs.harvard.edu/abs/2012ApJ...758..126S} {758, 126}

\bibitem[\protect\citeauthoryear{{Schekochihin} \& {Cowley}}{{Schekochihin} \&
  {Cowley}}{2007}]{schekochihin07}
{Schekochihin} A.~A.,  {Cowley} S.~C.,  2007, {Turbulence and Magnetic Fields
  in Astrophysical Plasmas}.
Springer, p.~85

\bibitem[\protect\citeauthoryear{{Schekochihin}, {Cowley}, {Kulsrud}, {Hammett}
   \& {Sharma}}{{Schekochihin} et~al.}{2005}]{schekochihin05}
{Schekochihin} A.~A.,  {Cowley} S.~C.,  {Kulsrud} R.~M.,  {Hammett} G.~W.,
  {Sharma} P.,  2005, \mn@doi [\apj] {10.1086/431202}, \href
  {http://adsabs.harvard.edu/abs/2005ApJ...629..139S} {629, 139}

\bibitem[\protect\citeauthoryear{{Shaw}, {Nagai}, {Bhattacharya}  \&
  {Lau}}{{Shaw} et~al.}{2010}]{shaw10}
{Shaw} L.~D.,  {Nagai} D.,  {Bhattacharya} S.,   {Lau} E.~T.,  2010, \mn@doi
  [\apj] {10.1088/0004-637X/725/2/1452}, \href
  {http://adsabs.harvard.edu/abs/2010ApJ...725.1452S} {725, 1452}

\bibitem[\protect\citeauthoryear{{Simionescu} et~al.,}{{Simionescu}
  et~al.}{2011}]{simionescu11}
{Simionescu} A.,  et~al., 2011, \mn@doi [Science] {10.1126/science.1200331},
  \href {http://adsabs.harvard.edu/abs/2011Sci...331.1576S} {331, 1576}

\bibitem[\protect\citeauthoryear{{Skillman}, {Hallman}, {O'Shea}, {Burns},
  {Smith}  \& {Turk}}{{Skillman} et~al.}{2011}]{skillman11}
{Skillman} S.~W.,  {Hallman} E.~J.,  {O'Shea} B.~W.,  {Burns} J.~O.,  {Smith}
  B.~D.,   {Turk} M.~J.,  2011, \mn@doi [\apj] {10.1088/0004-637X/735/2/96},
  \href {http://adsabs.harvard.edu/abs/2011ApJ...735...96S} {735, 96}

\bibitem[\protect\citeauthoryear{{Subramanian}, {Shukurov}  \&
  {Haugen}}{{Subramanian} et~al.}{2006}]{subramanian06}
{Subramanian} K.,  {Shukurov} A.,   {Haugen} N.~E.~L.,  2006, \mn@doi [\mnras]
  {10.1111/j.1365-2966.2006.09918.x}, \href
  {http://adsabs.harvard.edu/abs/2006MNRAS.366.1437S} {366, 1437}

\bibitem[\protect\citeauthoryear{{Turk}, {Smith}, {Oishi}, {Skory}, {Skillman},
  {Abel}  \& {Norman}}{{Turk} et~al.}{2010}]{turk2010yt}
{Turk} M.~J.,  {Smith} B.~D.,  {Oishi} J.~S.,  {Skory} S.,  {Skillman} S.~W.,
  {Abel} T.,   {Norman} M.~L.,  2010, {yt: A Multi-Code Analysis Toolkit for
  Astrophysical Simulation Data}, Astrophysics Source Code Library (\mn@eprint
  {ascl} {1011.022})

\bibitem[\protect\citeauthoryear{{Tzeferacos} et~al.,}{{Tzeferacos}
  et~al.}{2012}]{tzeferacosetal12}
{Tzeferacos} P.,  et~al., 2012, \mn@doi [High Energy Density Physics]
  {10.1016/j.hedp.2012.08.001}, \href
  {http://adsabs.harvard.edu/abs/2012HEDP....8..322T} {8, 322}

\bibitem[\protect\citeauthoryear{{Uchiyama}, {Aharonian}, {Tanaka}, {Takahashi}
   \& {Maeda}}{{Uchiyama} et~al.}{2007}]{uchiyama07}
{Uchiyama} Y.,  {Aharonian} F.~A.,  {Tanaka} T.,  {Takahashi} T.,   {Maeda} Y.,
   2007, \mn@doi [\nat] {10.1038/nature06210}, \href
  {http://adsabs.harvard.edu/abs/2007Natur.449..576U} {449, 576}

\bibitem[\protect\citeauthoryear{{Vink} \& {Laming}}{{Vink} \&
  {Laming}}{2003}]{vink03}
{Vink} J.,  {Laming} J.~M.,  2003, \mn@doi [\apj] {10.1086/345832}, \href
  {http://adsabs.harvard.edu/abs/2003ApJ...584..758V} {584, 758}

\bibitem[\protect\citeauthoryear{{Vogt} \& {En{\ss}lin}}{{Vogt} \&
  {En{\ss}lin}}{2005}]{Vogt05}
{Vogt} C.,  {En{\ss}lin} T.~A.,  2005, \mn@doi [\aap]
  {10.1051/0004-6361:20041839}, \href
  {http://adsabs.harvard.edu/abs/2005A%26A...434...67V} {434, 67}

\bibitem[\protect\citeauthoryear{{Wong} et~al.,}{{Wong} et~al.}{2008}]{wong08}
{Wong} T.,  et~al., 2008, \mn@doi [\mnras] {10.1111/j.1365-2966.2008.13107.x},
  \href {http://adsabs.harvard.edu/abs/2008MNRAS.386.1069W} {386, 1069}

\bibitem[\protect\citeauthoryear{{Yang}, {Ruszkowski}  \& {Zweibel}}{{Yang}
  et~al.}{2013}]{yang13}
{Yang} H.-Y.~K.,  {Ruszkowski} M.,   {Zweibel} E.,  2013, \mn@doi [\mnras]
  {10.1093/mnras/stt1772}, \href
  {http://adsabs.harvard.edu/abs/2013MNRAS.436.2734Y} {436, 2734}

\bibitem[\protect\citeauthoryear{{Zhuravleva} et~al.,}{{Zhuravleva}
  et~al.}{2013}]{zhuravleva13}
{Zhuravleva} I.,  et~al., 2013, \mn@doi [\mnras] {10.1093/mnras/stt1506}, \href
  {http://adsabs.harvard.edu/abs/2013MNRAS.435.3111Z} {435, 3111}

\bibitem[\protect\citeauthoryear{{Zhuravleva} et~al.,}{{Zhuravleva}
  et~al.}{2015}]{zhuravleva15}
{Zhuravleva} I.,  et~al., 2015, \mn@doi [\mnras] {10.1093/mnras/stv900}, \href
  {http://adsabs.harvard.edu/abs/2015MNRAS.450.4184Z} {450, 4184}

\bibitem[\protect\citeauthoryear{{van Weeren}, {R{\"o}ttgering}, {Br{\"u}ggen}
  \& {Hoeft}}{{van Weeren} et~al.}{2010}]{van-weeren10}
{van Weeren} R.~J.,  {R{\"o}ttgering} H.~J.~A.,  {Br{\"u}ggen} M.,   {Hoeft}
  M.,  2010, \mn@doi [Science] {10.1126/science.1194293}, \href
  {http://adsabs.harvard.edu/abs/2010Sci...330..347V} {330, 347}

\makeatother
\end{thebibliography}

\end{CJK}
\end{document}